\documentclass{aa}  
\usepackage{natbib}
\usepackage{graphicx}
\usepackage{txfonts}
\usepackage{longtable}
\usepackage{lscape}
\usepackage{placeins}
\begin{document}
\title{TIMASSS: The IRAS16293-2422 Millimeter And Submillimeter Spectral Survey \thanks{Based on observations with the IRAM 30\,m telescope (IRAM is supported by INSU/CNRS (France), MPG (Germany) and IGN (Spain)), and with the JCMT 15\,m telescope (operated by the Joint Astronomy Centre on behalf of the Particle Physics and Astronomy Research Council of the United Kingdom, the Netherlands Organisation of Scientific Research, and the National Research Council of Canada).}}
\subtitle{I. Observations, calibration and analysis of the line kinematics}
 \author{E.~Caux\inst{1,2} \and C.~Kahane\inst{3} \and A.~Castets\inst{4,5,3} 
 \and A.~Coutens\inst{1,2} \and C.~Ceccarelli\inst{3}  \and A.~Bacmann~\inst{4,5,3}
 \and S.~Bisshop~\inst{6,7} \and S.~Bottinelli\inst{1,2} \and C.~Comito\inst{8}
  \and F.P.~Helmich\inst{9} \and B.~Lefloch\inst{3} \and B.~Parise\inst{8} \and P.~Schilke\inst{8,10}
  \and A.G.G.M.~Tielens\inst{6,9,11} \and E.~van~Dishoeck\inst{6,12} \and C.~Vastel\inst{1,2} \and
  V.~Wakelam\inst{4,5} \and A.~Walters\inst{1,2}}
  
  

\institute{Universit\'e de Toulouse; UPS-OMP; IRAP; Toulouse, France, \email{caux@cesr.fr}
\and CNRS; IRAP; 9 Av. colonel Roche, BP 44346, F-31028 Toulouse cedex 4, France
\and Laboratoire d'Astrophysique de Grenoble, UMR 5571-CNRS, Universit\'e Joseph Fourier, Grenoble, France
\and Universit\'e de Bordeaux, Laboratoire d'Astrophysique de Bordeaux, 33000 Bordeaux, France 
\and CNRS/INSU, UMR 5804, B.P. 89, 33271 Floirac cedex, France
\and Leiden Observatory, Leiden University, P.O. Box 9513, NL 2300 RA Leiden, The Netherlands
\and Center for Star and Planet Formation, University of Copenhagen, Oster Voldgade 5-7, DK-1350, Copenhagen, Denmark
\and Max-Planck-Institut f\"ur Radioastronomie, Auf dem H\"urgel 69, 53121 Bonn, Germany
\and SRON Netherlands Institute for Space Research, PO Box 800, 9700 AV, Groningen, The Netherlands
\and I. Physikalisches Institut, Universit\"at zu Kf\"oln, Z\"ulpicher Str. 77, 50937 K\"oln, Germany
\and Kapteyn Astronomical Institute, University of Groningen, PO box 800, 9700 AV Groningen, Netherlands
\and Max-Planck Institute f\"ur Extraterrestrische Physik,Giessenbachstr. 1, D-85748 Garching, Germany.}

\date{Received xxx, xxxx; accepted xxx, xxx}


  \abstract
  {Unbiased spectral surveys are powerful tools to study the
    chemistry and the physics of star forming regions, as they can give a complete census of the molecular content and the observed
    lines probe the physical structure of the source.}
  {While unbiased surveys in the millimeter and sub-millimeter ranges
    observable from ground-based telescopes have previously been obtained
    towards several high mass protostars, very little exists on low
    mass protostars, believed to resemble our own Sun's progenitor. Aiming
    to fill up this gap, we carried out a complete spectral survey of
    the bands at 3, 2, 1 and 0.8\,mm towards the solar type protostar
    IRAS16293-2422.  }
  {The observations covered about 200\,GHz and were obtained with the
    IRAM-30\,m and JCMT-15\,m telescopes during about 300 hours of
    observations.  Particular attention was devoted to the
    inter-calibration of the obtained spectra with previous
    observations.  All the lines detected with more than 3\,$\sigma$ and free from obvious blending effects
    were fitted with Gaussians to estimate their basic kinematic
    properties. }
  {More than 4000 lines were detected (with $\sigma \geq $\,3) and identified, yielding
    a line density of approximatively 20 lines per GHz, comparable to
    previous surveys in massive hot cores. The vast majority ($\sim 2/3$) of
    the lines are weak and due to complex organic molecules. The
    analysis of the profiles of more than 1000 lines belonging 70 species firmly establishes the presence of two distinct velocity
    components, associated with the two objects, A and B, forming the IRAS16293-2422 binary system. In the source
    A, the line widths of several species increase with the upper level energy of the transition, a behavior compatible with gas
    infalling towards a $\sim 1$\,M$_\odot$ object. The source B, which does not show this effect, might have a much lower 
    central mass of $\sim 0.1$\,M$_\odot$. The difference in the rest velocities of both objects is consistent with the hypothesis that 
    the source B rotates around the source A. }
  {This spectral survey, although obtained with single-dish telescope with a low spatial resolution, allows to separate the emission from 2 different components, thanks to the large number of lines detected. The data of the survey are public and can be retrieved on the web site http://www-laog.obs.ujf-grenoble.fr/heberges/timasss.}


    \titlerunning{IRAS16293-2422 spectral survey}
    \authorrunning{Caux et al.}

  \maketitle
%

\section{Introduction}
It is well known that the chemical composition of the gas from which
the star forms influences the process of the star formation and it is,
in turn, influenced by it. The first obvious example is that the Jeans
mass depends on the gas temperature which, in turn, is set by the
molecular line cooling in a large range of densities and temperatures.
The chemical composition of the gas is here of paramount importance,
as the cooling is indeed dominated by different species as a function
of the gas temperature and density and the elemental abundance
\citep{2001ApJ...557..736G}.  A second classical example is the slow
contraction of the prestellar cores which is governed by ambipolar
diffusion. Since only ions feel the magnetic field which counteracts
the gravitational force, the chemical composition of the gas, which
determines the ions abundance, is crucial. In addition to this, since
the gas chemical composition is largely affected by the star formation
process, its study in star forming regions is a powerful diagnostic
tool to track down the various processes at work. Finally, the study
of the chemical composition in regions forming solar type stars is of
particular importance, as it helps to understand the history of the
formation of our own Solar System. For example, the comparison of the
chemical composition of comets with that of solar type protostars or
protoplanetary disks is used to trace back the origin of the former
\citep{2004IAUS..202..178C}.  The same applies to the studies of the
molecular content of meteorites. In particular, the recent claim that
the amino acids found in carbonaceous meteorites may have formed
during the first phases of the life of the Solar System is based on
the measured large deuteration in the meteoritic amino acids and in
the protostellar environment \citep{2005GeCoA..69..599P}.  In this
context, unbiased spectral surveys in millimeter and submillimeter
wavelengths are particularly relevant as they allow to detect heavy
and large molecules, and, specifically, complex organic molecules.  In
summary, unbiased spectral surveys in the millimeter to submillimeter
wavelengths are a powerful method to characterise the molecular
content of astrophysical objects, and the only way to obtain a
complete census of the chemical species.

There are at least two other aspects that make unbiased spectral
surveys precious tools for studying the star formation process. First,
in general they provide multiple lines from the same molecule,
allowing multi-frequency analysis and modelling. Since different lines
from transitions with different upper level energies and Einstein
coefficients are excited at different temperatures and densities, they
probe different regions in the line of sight. A careful analysis can,
therefore, distinguish between the various physical components in the
beam. If one adds also the kinematic information provided by the line
profiles, the method can be so powerful that it can identify
sub-structures in the line of sight, even if the spatial resolution of
the observations is inadequate. In the present article, we provide an
example of this capability of unbiased spectral surveys.

Given their powerful diagnostic ability, several unbiased spectral
surveys in the millimeter and sub-millimeter bands accessible from
ground have been obtained in the past in the direction of star forming
regions. A complete list of these surveys can be found in
\cite{2009ARA&A..47..427H}.  By far the most targeted sources are hot
cores, the regions of high mass protostars formation, where the dust
temperature exceeds the sublimation temperature of the water-ice grain
mantles, $\sim 100$\,K.  The combined effect of the mantle sublimation
and the high gas temperature triggers a singular and rich
chemistry. At the same time, the relatively high densities ($\geq
107$\,cm$^{-3}$) and temperatures ($\geq$\,100\,K) are favorable for
the excitation of several high lying transitions. The result is that
extremely rich spectra are observed towards the hot cores. About a
dozen hot cores have been targeted in different bands
(\citealt{1997ApJS..108..301S, 1997A&AS..124..205H,
  1998A&AS..133...29H, 2001ApJS..132..281S, 2010A&A...517A..96T}). One of the most studied hot cores is
the Orion-KL source. Spectral surveys covering almost all the bands
accessible from the ground have been obtained, from about 80 to 900
GHz, detecting thousands of lines from hundreds of species and
relative isotopologues (see for instance \citet{1997ApJS..108..301S,
  2001ApJ...551..333L, 2005ApJS..156..127C, 2007A&A...476..791O} and
references therein;  \citet{2007A&A...466..255D, 2009A&A...500.1109C, 2009A&A...493..565M, 2010A&A...517A..96T}. 
In addition, the
500--2000\,GHz range is observed with the HIFI spectrometer
 \citep{2010A&A...518L...6D} on board the recently launched Herschel
 \footnote{Herschel is an ESA
  space observatory with science instruments provided by European-led
  principal Investigator consortia and with important participation
  from NASA} satellite \citep{2010A&A...518L...1P}, in the Key Program
HEXOS ({\it http://www.submm.caltech.edu/hexos/}).  Similarly, the
500---2000\,GHz range is observed in other hot cores, in the
Key Program CHESS ({\it
  http://www-laog.obs.ujf-grenoble.fr/heberges/chess/}). Preliminary
results of the surveys performed in these two Herschel Key Programs
can be found in \citep{2010A&A...521L..20B} and
\citep{2010A&A...521L..22C}, respectively.

Although less massive and less luminous, solar type protostars also
possess regions where the dust mantles sublimate, yielding similar properties as
those of hot cores (see \citet{2007msl..confE...1C} and references
therein). These regions have been baptized hot corinos, to make it clear
that they share similarities with hot cores but are not just scaled
version of them (see also \citealt{2007A&A...463..601B}). The interest in
observing hot corinos, whose sizes are comparable to the Solar System
sizes, is amplified by the fact that they likely resemble the Solar
Nebula. In other words, their study corresponds to an archeological
study of our ancestor Solar System.

So far, only one (partial) spectral survey has been obtained towards a
solar type protostar \citep{1994ApJ...428..680B,
  1995ApJ...447..760V}. The targeted source was IRAS16293-2422
(hereinafter IRAS16293), in the L1689N cloud (d\,=\,120\,pc,
\citealt{2008ApJ...675L..29L}). This survey partially covered the two
windows in the 200\,GHz and 350\,GHz bands accessible from ground, and
has been obtained with the JCMT and CSO telescopes. The sensitivity
achieved ($\sim 40$ mK ) allowed the detection of 265 lines from 24
species, namely the most abundant molecules: CO, H$_2$CO, CH$_3$OH,
SO, SO$_2$ etc... 
Later, more sensitive observations have shown that the IRAS16293
  line spectrum is rich in 
complex organic molecules \citep{2000A&A...357L...9C,
  2003ApJ...593L..51C}, and doubly \citep{1998A&A...338L..43C} and triply
\citep{2004A&A...416..159P} deuterated molecules. In
retrospective, further support for the interest in obtaining an
unbiased spectral survey towards IRAS16293 is provided by the recent
Herschel/HIFI data in the 555---636\,GHz range: it shows that, while
IRAS16293 has much less lines than the $2\times106$ L$_\odot$ source
NGC6334I, the same number of species is detected in both sources
\citep{2010A&A...521L..22C}.

Several studies have been carried out towards IRAS16293, both with
single dish telescopes and interferometers. The emerging overall
picture is that IRAS16293 is a protobinary system
\citep{1989ApJ...337..858W, 1992ApJ...385..306M} surrounded by an
envelope of about 2\,M$_\odot$ \citep{2010A&A...519A..65C}. The
structure of the envelope has been the target of several studies
\citep{2000A&A...355.1129C, 2002A&A...390.1001S, 2005ApJ...631L..77J}.
The most recent one by \cite{2010A&A...519A..65C} concludes that the
envelope density follows a r$^{-2}$ power law at distances larger than
about 1300\,AU and r$^{-3/2}$ innerwards. The grain mantles are
predicted to sublimate at a distance of 75\,AU, where the density is
equal to $2\times108$\,cm$^{-3}$. The envelope extends from about 25
AU to about 6000 AU from the center. Inside the envelope, the two
sources, A (South-East) and B (North-West), of the binary system are
separated by about 4$"$ (separation measured from interferometer 
observations at a spatial resolution of about 1$"$), equivalent to a linear distance of 480\,AU.
The source B is brighter than the source A in the millimeter continuum and
in several ``cold envelope'' molecular lines, whereas the source A seems
to be brighter in several hot corino like molecular lines
\citep{2004ApJ...616L..27K, 2004ApJ...617L..69B,
  2005ApJ...632..371C}. Finally, \citet{2005ApJ...632..371C} have
claimed that source A might be a multiple system and very recent
observations \citep{2010ApJ...712.1403P} suggest that A is itself a
binary system of 0.5 and 1.5\,M$_\odot$ respectively.

Despite its relatively complex structure at arcsec scales, IRAS16293
remains the brightest and best source to carry out a detailed study of
the gas chemical composition in the first phases of the formation of a
solar type star. As discussed above, the best way for that is to
obtain unbiased spectral surveys, as sensitive as possible. In this
paper, we present the results of the most sensitive unbiased spectral
survey of the bands between 80 and 366\,GHz observable from ground
based telescopes obtained so far in direction of IRAS16293.  This
study is part of a more general project that plans on observing also
the 500-2000\,GHz frequency range with the spectrometer HIFI on board
the Herschel satellite, in the context of the Herschel Key Program
CHESS ({\it Chemical Herschel Surveys of Star forming regions:
  http://www-laog.obs.ujf-grenoble.fr/heberges/chess/}, \citep{2010A&A...521L..22C}.

\section{Observations}

The observations were obtained at the IRAM-30\,m (frequency range 80-280\,GHz) 
and JCMT-15\,m (frequency range 328-366\,GHz) telescopes during the
period January 2004 to August 2006. Overall, the observations required
a total of about 300 hours ($\sim$ 200\,hr at IRAM and $\sim$\,100\,hr at
JCMT) of observing time.  The beam of the survey varied between
9{\arcsec} and 33{\arcsec}, depending on the used telescope and
frequency, and the spectral resolution ranged between 0.3 and
1.25\,MHz, corresponding to velocity resolutions between 0.51 and 2.25\,km\,s$^{-1}$. 
The achieved rms varied between 4 and 14\,mK in
1.5\,km\,s$^{-1}$ bins. The observations were centered on the B
(North-West) component at $\alpha$(2000.0)\,=\,16$^h$ 32$^m$ 22.6$^m$,
$\delta$(2000.0)\,=\,$-$24$^{\circ}$ 28$'$ 33$'$. 
A and B components, separated by 4{\arcsec}, are both inside the beam of our
observations at all frequencies. 
However, at the highest frequencies observed with the IRAM 30\,m telescope (i.e. the 1\,mm band), 
the attenuation of emission from source A is not negligible as will be discussed in section 5.
All observations were performed in Double-Beam-Switch observing mode,
with a 90{\arcsec} throw.  Pointing and focus were checked every two
hours on planets and on the continuum radio sources 1741-038 and 1730-130.  
Table \ref{tab-observations} summarizes the observed
bands and the details of the observations. More details are given
below for the IRAM-30\,m and JCMT-15\,m telescope observations,
respectively.  Because of the different weather conditions during the
different runs, the system temperatures varied largely. However,
during the data processing, scans with too large system temperatures
were removed before averaging.

%
%
\begin{table*}[htb]
\caption{Parameters of the observations at IRAM-30\,m and JCMT-15\,m telescopes.}
\begin{center}
\begin{tabular}{cccccccccccc}
\hline \hline\\[-1.8ex]
Telescope & Frequency & \multicolumn{2}{c}{Resolution} & $^{(1)}$Backend & $^{(2)}$rms & Tsys & HPBW & Beam & $^{(3)}\sigma_{cal}$ & $^{(4)}$N$_{cal}$ & $^{(5)}$P$_{cal}$ \\
 &(GHz) &(MHz) &(km.s$^{-1}$) & &(mK) &(K) & (arcsec) & efficiency &(\%) & lines &(\%)\\[0.5ex]
\hline\\[-1.8ex]
IRAM &80 - 115.5 &0.32 &0.81-1.17 &VESPA &2-8 &90-400 & 30-21 &0.80-0.78 &11 &28 &15 \\
IRAM &129 - 177 &0.32 &0.53-0.72 &VESPA &5-14 &200-1000 &19-14 &0.76-0.69 &17 & 22 & 12 \\
IRAM & &1.0 &1.65-2.25 &1\,MHz FB & & & & & 10$^{(6)}$ & 84$^{(6)}$ & 95$^{(6)}$ \\ 
IRAM &197 - 265 &1.0 &1.13-1.52 &1\,MHz FB &7-13 &180-1200 & 12-9 &0.65-0.51 &17 &36 &10 \\
IRAM & &1.25 &1.41-1.90 &VESPA  & & & & & & &\\
IRAM &265 - 280 &1.25 &1.34-1.41 &VESPA &9-17 &470-4200 & 9 &0.51-0.47\\
JCMT &328 - 366 &0.625 &0.51-0.57 &ACSIS &4-9 &90-400 & 14 &0.56-0.53 &18 & 26 &25\\[0.5ex]
\hline
\end{tabular}
\end{center}
$^{(1)}$In the 2mm band, each frequency setting is observed twice, i.e. once at 
each of the two spectral resolutions (0.32 and 1\,MHz), while in the
197-265\,GHz band, each frequency setting is observed only one time, at
a slightly different spectral resolution (1 or 1.25\,MHz), $^{(2)}$rms
is given in 1.5\,km.s$^{-1}$ bins, $^{(3)}\sigma_{Cal}$ is the
calibration uncertainty, $^{(4)}$N$_{cal}$ is the number of compared
lines for calibration purposes, $^{(5)}$P$_{cal}$ is the percentage of
compared spectra for calibration purposes, $^{(6)}$these values refer
to the ''internal'' comparison between VESPA and 1\,MHz FB spectra
simultaneously observed; the other ones refer to "external"
comparisons with published or previously obtained
observations. 
\label{tab-observations}
\end{table*}

\subsection{IRAM Observations}\label{iram-observations}

The following three bands were almost fully covered by observations at the
IRAM-30\,m telescope: 3\,mm band (80-115.5\,GHz), 2\,mm band
(129-177.5\,GHz) and 1~mm band (198-281.5\,GHz).

In all IRAM-30\,m observations, two frequency ranges were observed simultaneously, 
with two SIS receivers with orthogonal polarizations for each frequency range, 
in the following configuration:  3\,mm receivers (A100 \& B100) in parallel with 1\,mm receivers (A230 \& B230),
and 2\,mm receivers (C150 \& D150)  in parallel  with 1\,mm receivers (C270 \& D270). 
Because of the limitation (at the time of the observations) of the IRAM-30\,m backend capabilities in
terms of instantaneous frequency bandwidth and spectral resolution, we
privileged the largest possible spectral bandwidth to cover the
IRAM-30\,m bands in the smallest observing time. 
For simultaneous observations in the 3\,mm and 1\,mm bands, the VESPA autocorrelator
was split in 4 parts, 2 of them covering the whole IF band of the
A\&B100 receivers (0.5\,GHz) with 320\,kHz spectral resolution and the
two others covering half of the IF band of the A\&B230 receivers
(1\,GHz) with 1250\,kHz spectral resolution. The second half of the IF
band of the A\&B230 receivers was covered with the 1\,MHz Filter Banks
(FB). For simultaneous observations in the 2\,mm and upper 1\,mm
bands, the VESPA autocorrelator was split in 4 parts, 2 of them
covering half of the IF band of the C\&D150 receivers (0.5\,GHz) with
320\,kHz sampling two others covering half of the IF band of the
C270\&D270 receivers (1\,GHz) with 1250\,kHz sampling. The second half
of the IF band of the C\&D150 receivers was covered with the 1\,MHz
FB.

The configuration for observations in the 2\,mm band resulted in
different spectral resolutions for each half of the IF band of the
receivers (320\,kHz $\sim$ 0.65\,km\,s$^{-1}$ and 1\,MHz
$\sim$ 2\,km\,s$^{-1}$).  Therefore, we shifted the tuning frequency of
the receivers by only 0.5\,GHz from one tuning to the next one to
cover the entire 2\,mm band at the highest and at the smallest spectral
resolution respectively. As a consequence, two different datasets were
obtained, one at high resolution (generally used for studying
brighter lines), and one at low resolution (for the faint lines).

\subsection{JCMT Observations}\label{jcmt-observations}
The JCMT-15\,m observations covered the 328 to 366\,GHz frequency range. They
were obtained with 345\,GHz SIS receiver RxB3 in
dual-channel single-sideband (SSB) mode. Each polarization of the
receiver was connected to a unit of the ACSIS autocorrelator providing
a bandwidth of 0.5\,GHz for a spectral resolution of 625\,kHz. At 345\,GHz, 
this yields a velocity resolution of about 0.5\,km\,s$^{-1}$. 

%
\section{Calibration}
\subsection{Method}
At the 30\,m telescope, the calibration was performed with a cold and a warm
absorbers, and the atmospheric opacity was obtained using the ATM program (Cernicharo, 1985, IRAM internal report). 
At the JCMT, line strengths were calibrated via the chopper wheel method \citep{1981ApJ...250..341K}.

Our spectral survey does not allow to estimate the calibration uncertainties from line observation redundancy: 
each spectral range has been observed only once and there is only negligible frequency overlap between adjacent spectra. 
From our simultaneous observations with two receivers in the 1\,mm range we may estimate the receivers contribution, but 
to derive the total calibration uncertainties of the survey, we have performed a detailed comparison between our 
observations and previous observations. As our comparisons rely only on observations obtained with the same telescopes 
towards the same position (namely source B), there is no bias due to different sources dilution in the beams and our results 
are not affected by the underestimate of source A contribution at high frequency. 
The comparison includes virtually all
the published data towards IRAS16293 obtained with the IRAM-30\,m and
the JCMT-15\,m telescopes, as well as unpublished data obtained with the
IRAM-30\,m telescope.  The list of the articles used for this comparison
is the following: {\it a) IRAM-30\,m bands:} \citet{1998A&A...338L..43C,
  2000A&A...359.1169L, 2003ApJ...593L..51C, 2004A&A...413..609W,
  2002A&A...393L..49P, 2005A&A...441..171P}; {\it b) JCMT-15\,m band:}
\citet{1994ApJ...428..680B, 1995ApJ...447..760V, 2000A&A...359.1169L,
  2002A&A...390.1001S, 2004A&A...416..159P}.  
%
%
Table \ref{tab-observations} reports the percentage of the survey spectra
that could be cross-checked and calibrated against previously
published data for each band. In addition, we cross-checked the
calibration in the 2\,mm band by comparing the data obtained with
1\,MHz and 320\,kHz resolution.  Finally, we estimated the calibration
uncertainty due to the receivers by comparing lines in the 1\,mm band
observed with the two receivers A230 and B230.

To quantify the differences, we obtained Gaussian fits of the
considered lines and compared their characteristics 
(integrated intensity, peak intensity and full width at half maximum,
FWHM), with the previously published values.  Note that the comparison
was performed in the main beam brightness scale (T$_{mb}$), based on
the T$_A^*$ /T$_{mb}$ beam efficiency factors given in Table
\ref{tab-observations}. This method provides two types of check: i)
the average uncertainty over each band; and ii) possible specific
calibration problems on single settings.

\subsection{Calibration uncertainties}
With the method described above to quantify the calibration
uncertainty in the survey, we obtained the following results.\\

\noindent
{\it FWHM:} \\
In all the frequency ranges, except 1mm, the agreement between the
survey and the published FWHM values is within 15-20\% and no
systematic trend is observed. In contrast, in the 1\,mm band the
FWHM of the survey lines appears systematically broader by $\simeq$
1\,kms$^{-1}$ than the published values. This is likely due to the
relatively poorer spectral resolution of our survey in this range (0.8
to 1.9\,kms$^{-1}$) compared to the linewidths (on average $\simeq$
4-5\,kms$^{-1}$).
\\

\noindent
{\it Integrated and peak intensity:}\\
The comparison of the integrated and peak intensity of the lines of
the survey with published values yields the same results, when the
difference due to the spectral resolution described above is taken
into account.  Besides, since the derivation of the integrated
intensities does not depend on the line shape, we choose to quantify
the calibration uncertainties by comparing the integrated line
intensities.
\\

\begin{figure}[ht]
\includegraphics[angle=-0, width=1.0\linewidth]{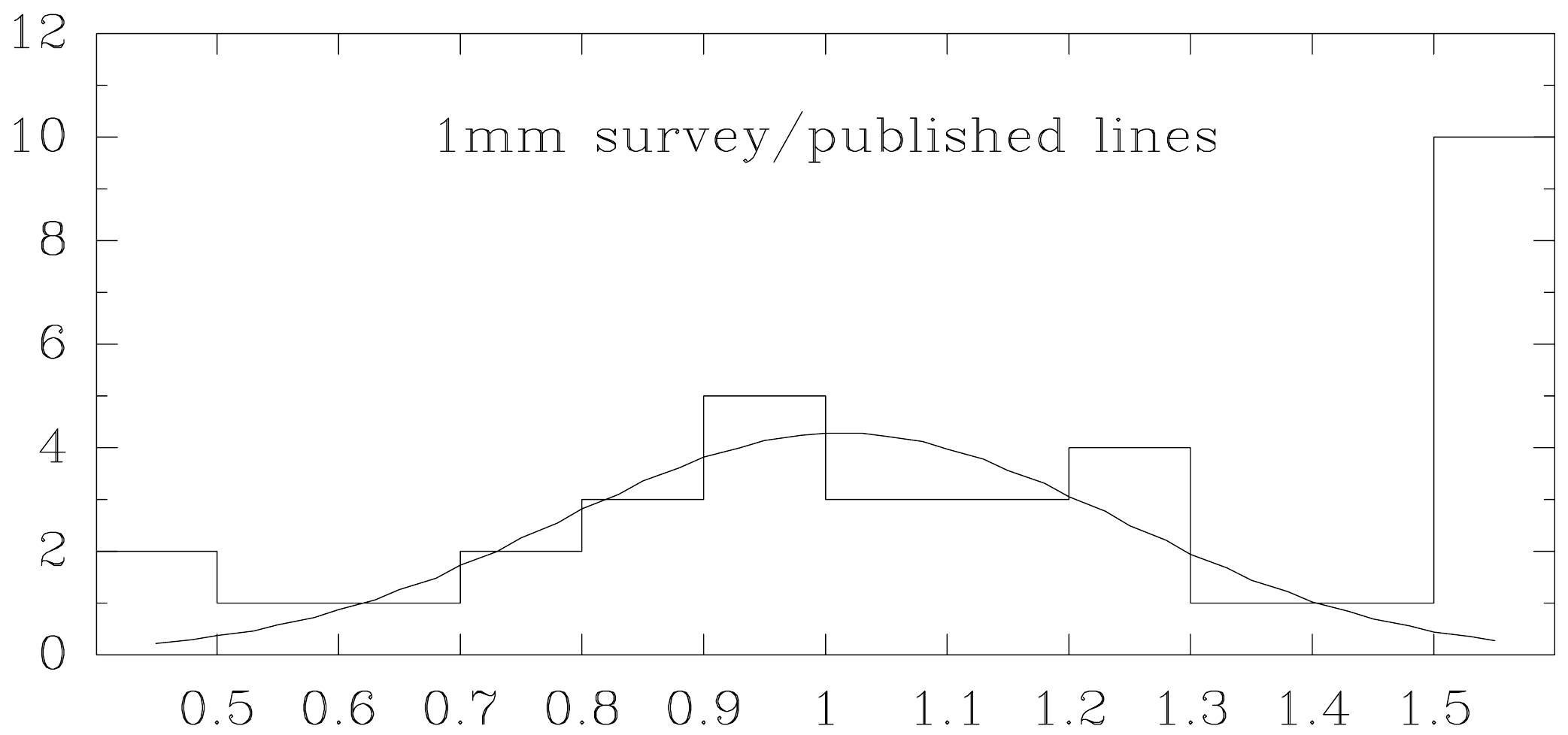}
\caption{Distribution of integrated intensity ratios of this survey's
  1\,mm lines compared to published observations obtained with the
  IRAM 30\,m telescope. The curve is a Gaussian fit to the histogram
  ruling out the ``anomalous'' ratios ($\leq\,0.5$ or $\geq\,1.5$). It
  can be noticed that the 12 ``anomalous'' ratios correspond to only 7
  ``anomalous spectra'' among the 165 spectra observed in the 1\,mm
  range.}
\label{1mm-distribution}
\end{figure}

\noindent
{\it Derived uncertainties:}\\ 
We can derive an estimate of the calibration uncertainty from the distribution of
the integrated intensity ratios. Fig~\ref{1mm-distribution} shows the
result for the 1mm band. If one excludes the two extremes at 
 $\leq\,0.5$ and $\geq\,1.5$ which correspond to spectra with ``anomalous intensity", 
the ratios distribution can be fitted by a Gaussian  with a mean value $R$, very close to 1 and 
a standard deviation $\sigma_{ratio}$. 
With the assumption that the relative uncertainties on the published intensities, 
$\sigma_{pub}$, and on the survey line intensities, $\sigma_{cal}$, are independent variables, 
the error propagation formula leads to the following relationship:
\begin{equation}
\sigma_{ratio}^2 = R^2 ( \sigma_{cal}^2 +  \sigma_{pub}^2)
\end{equation}
Most publications report calibration uncertainties of 15\% (or do not report any estimate). 
Except in the 3\,mm frequency range, where our comparisons suggest that the calibration 
uncertainty is probably somewhat lower than 15\%, we obtain consistent uncertainties either 
assuming $ \sigma_{pub}$ = 15\% for all the frequency ranges or assuming that our observations 
are representative of average observation conditions in each frequency range, i.e.  
$\sigma_{pub} = \sigma_{cal}$ and thus $\sigma_{cal}^2 \simeq \sigma_{ratio}^2 / 2$. 
Table \ref{tab-observations} reports the resulting calibration uncertainties for each observed band. 

It can be noticed that at 2\,mm, the ``external''
comparisons with published spectra, which include all uncertainty factors, lead to higher values 
than ``internal'' comparisons between the VESPA and 1\,MHz FB simultaneous observations, 
which take into account only the contribution of the backends. Similarly, at 1\,mm, comparison
of the line intensities observed simultaneously with the two receivers A230 and B230
shows that the receivers' contribution to the calibration uncertainties
is of the order of 10\%, whereas the total calibration uncertainty is
17\%.

Finally, the full list of lines used for the calibration comparison is
reported in the On Line Material, in Table \ref{tab_external_calib}
for a comparison between the survey lines and previous observations
obtained with the same telescopes and in Table \ref{tab_internal_calib} 
for a comparison between the survey lines observed simultaneously 
with VESPA and the 1\,MHz Filter Bank at the IRAM 30\,m telescope.


\addtocounter{table}{1}

\addtocounter{table}{1}

\section{Data release}

\begin{figure*}[!ht]
\centering
\includegraphics[angle=0,width=.9125\linewidth]{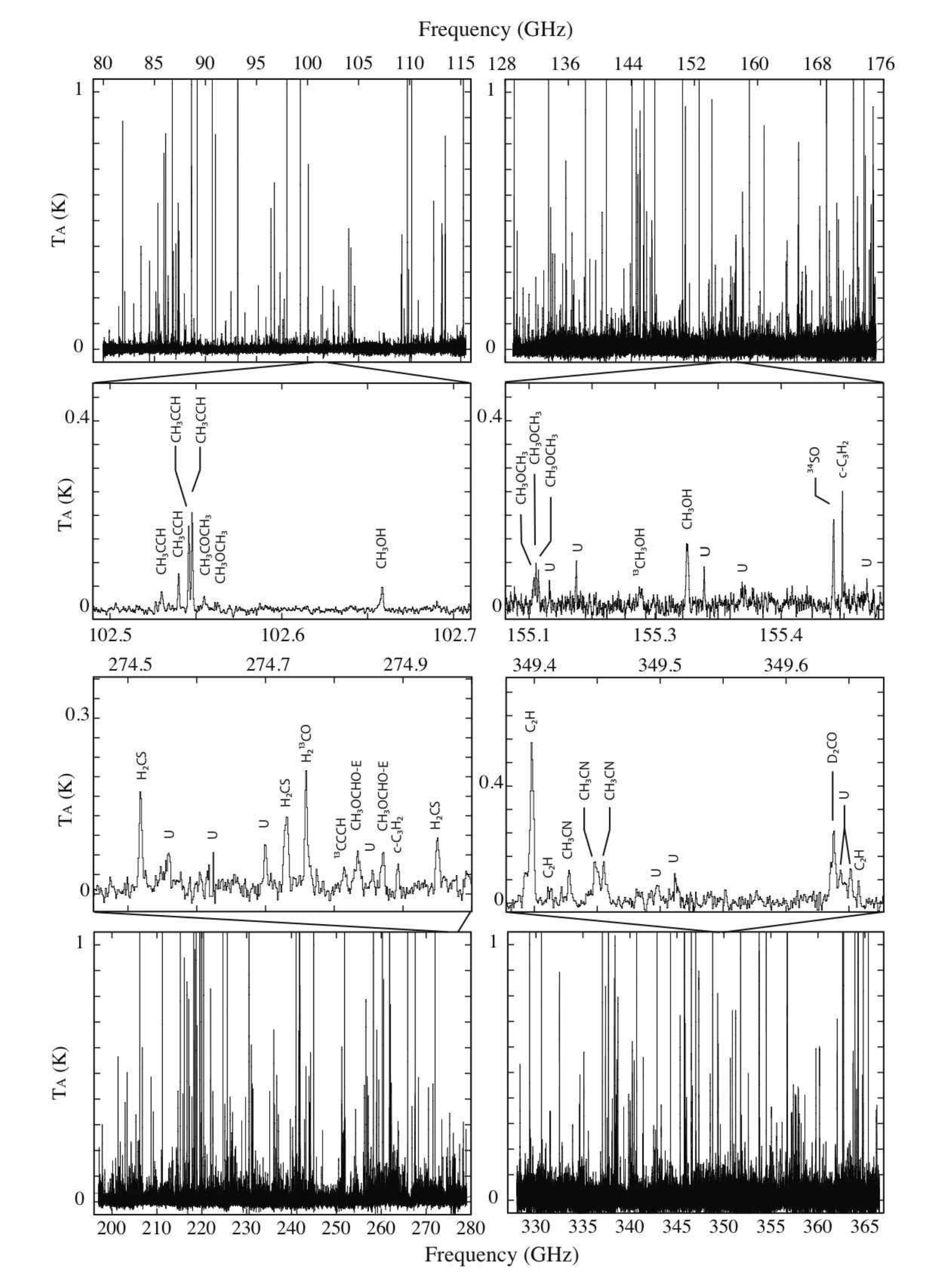}

\caption{The IRAS16293 spectra in the four bands of the survey. Upper
  panels: IRAM-30\,m 3\,mm and 2\,mm bands. Lower panels: IRAM-30\,m
  1\,mm and JCMT-15\,m 0.8\,mm bands. The middle panels are blow-ups of
  sample frequency ranges in the four bands respectively. These panels
  include lines identification based on the publicly available spectral
  databases (see text for details).}
\label{survey}
\end{figure*}
The data are made publicly available on the web site {\it
  http://www-laog.obs.ujf-grenoble.fr/heberges/timasss}.  The site
provides the files with the IRAM-30\,m and JCMT-15\,m data in CLASS 
format (http://www.iram.fr/IRAMFR/GILDAS). The intensities are in T${_A}
{^*}$.  Based on the discussion of the previous section, the potential
user is highly recommended to verify the calibration uncertainty of
the data that she/he wants to use by looking at Tables\,\ref{tab_external_calib} 
and \ref{tab_internal_calib}. We
emphasize that we did not apply any ``rescaling'' factor to the survey
data because the difference may be caused by a wrong calibration of
the published data rather than a wrong calibration of the survey
data. Only a very careful scientific analysis can assess what is the
best. It is, therefore, the user's responsibility to verify that the
data are correctly used, based on the information provided on the web
site.

\section{Results}\label{sec:results}

\subsection{Overall survey}\label{sec:overall-survey}

Figure \ref{survey} shows the full survey in the four bands and the
richness of the IRAS16293 line spectrum. On average about 20 lines per
GHz have been detected with a S/N larger than 3 in the 220\,GHz
frequency range covered by the survey. The line density seems to
slightly increase with frequency: 17 lines/GHz in the 3\,mm band, 19
and 23 lines/GHz in the 2\,mm and 1\,mm ranges respectively and up to 26
lines/GHz in the 0.9\,mm range. 

\begin{figure}[htb]
\centering
\includegraphics[angle=0, width=1.0\linewidth]{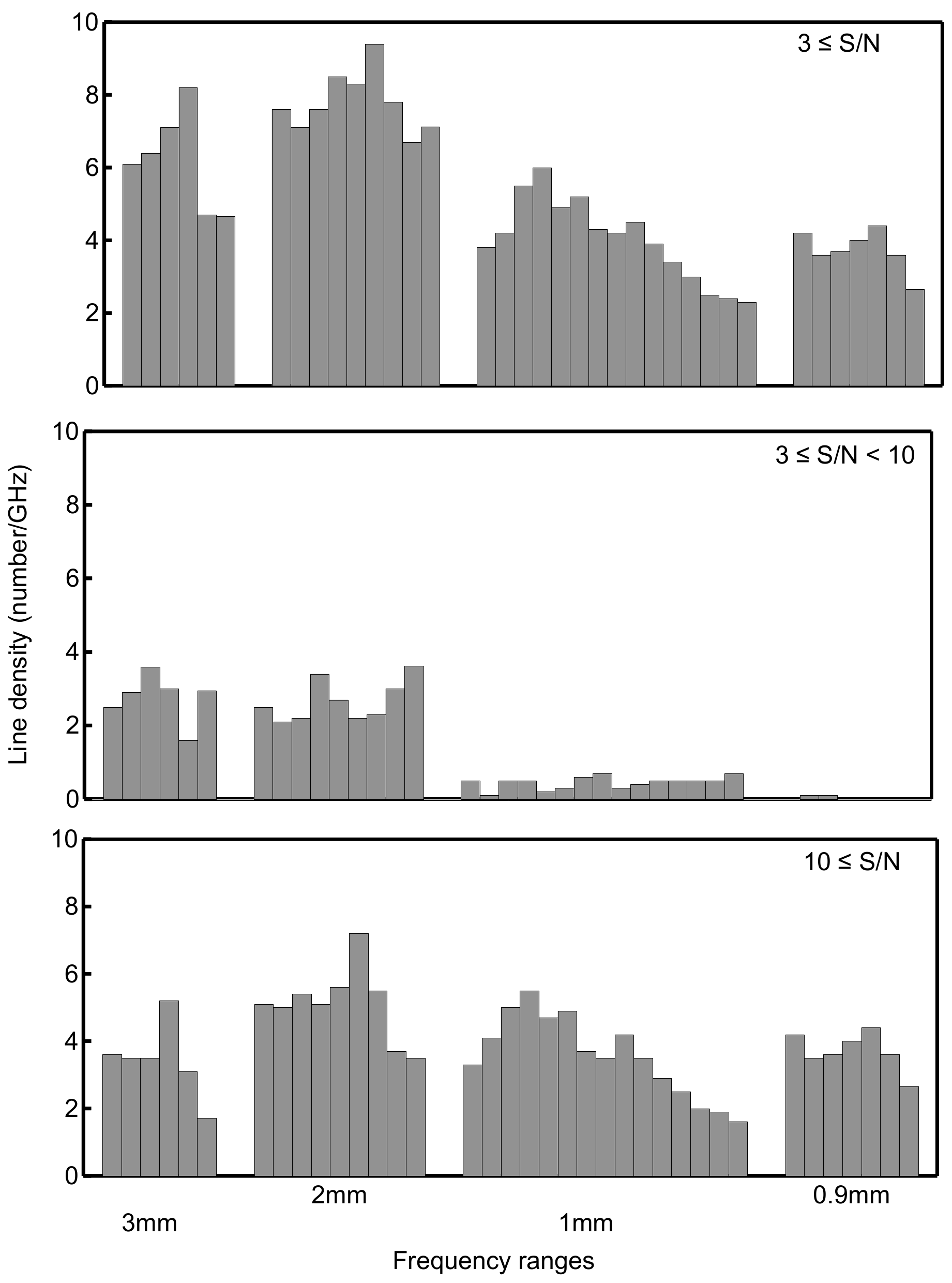}
\caption{Distribution of the density (averaged over 10\,GHz intervals)
  of identified lines (see text) in each of the four frequency
  ranges. The upper panel corresponds to lines with S/N
  $\leqslant$\,3, the central panel is restricted to lines with
  3\,$\leqslant$\,S/N\,$<$\,10 and the lower panel corresponds to
  lines with signal to noise ratios $\geqslant$\,10.}
\label{density}
\end{figure}

To quantify more rigorously the lines and species detected in the survey, we made Gaussian fits and line 
identification using the CASSIS package ({\it http://cassis.cesr.fr}\footnote{CASSIS has been developed by CESR-UPS/CNRS}.). 
The spectroscopic data come from the CDMS and JPL databases (\citealt{2001A&A...370L..49M, 2005JMoSt.742..215M, 1998JQSRT..60..883P} 
and references quoted on the databases to data producers for each species). In a few cases (ortho and para 
H$_2$CO for instance) a specific database with each form separated has been used (see http://cassis.cesr.fr).
For the D-bearing isotopologues of methanol, only the lines reported by \citep{2002A&A...393L..49P,  2004A&A...416..159P} 
are included in this paper.

Hereinafter we will only consider lines identified according to the following criteria:
i) lines belonging to species included in the JPL and CDMS databases or to the D-bearing isotopologues of methanol, 
ii) lines detected with more than 3$\sigma$ in the integrated line
intensity, iii) unblended lines and iv) lines with upper level energies
E$_{\rm up}$ lower than 250\,K. This last condition only limits the number
of methanol lines in this analysis, since such lines of other molecules are too weak
in any case. When applying these criteria, we end up with $\sim$\,1000
lines listed in Table \ref{lines-on-line} (Online Material). In the
table we report the line identification together with the result of
the Gaussian fit of each line (see also \S \ref{sec:line-kinematics}).

\addtocounter{table}{1}

\begin{figure}[htbc]
\centering
\includegraphics[angle=90, width=0.95\linewidth]{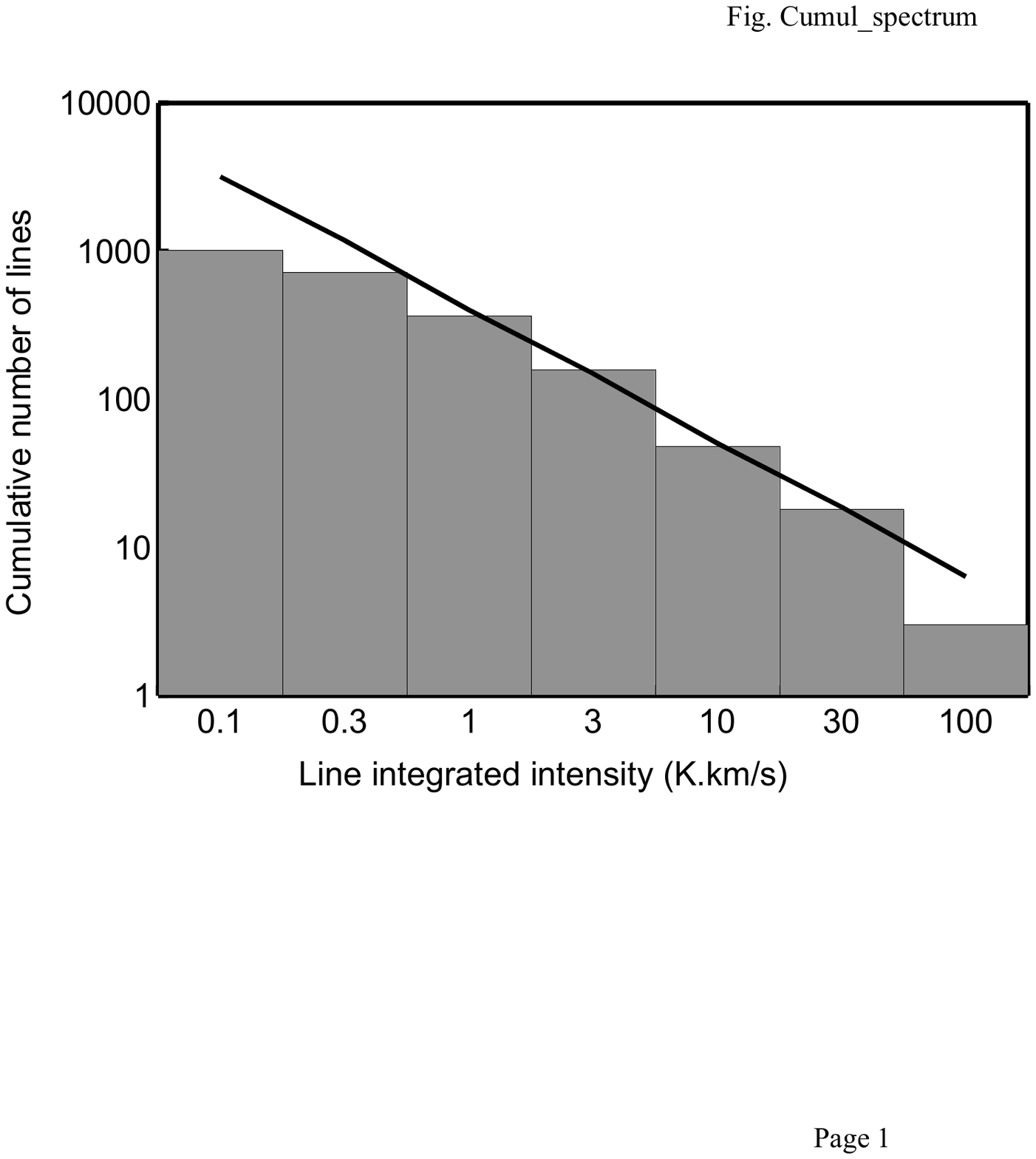}
\caption{Number of lines showing an integrated intensity
  larger than a given threshold. The solid line shows the single index
  power law best fit of the distribution.}
\label{cumul}
\end{figure}

Figure \ref{density} shows the line densities, limited to the lines
satisfying the above criteria, in each of the four survey frequency
ranges, for various signal to noise ratios. In the 3\,mm and 2\,mm
ranges, theses densities are a factor 2 to 3 smaller than the
estimates of the total densities including blended lines. This effect
is still stronger in the 1\,mm and 0.9\,mm ranges, where the lack of
weak lines ($3\,\leqslant\,S/N\,<\,10$) is particularly striking in
Fig. \ref{density}. This is a bias due to our selection criterion of
non-blended lines. In fact, these frequency ranges are rich in lines
from large molecules, which emit many weak lines, so that our
unblended line criterium filters out a large fraction of these moderate
signal to noise ratio lines. In contrast, the 2\,mm range, which
benefits from a better spectral resolution compared to the 3\,mm and
1\,mm ranges, suffers less from this selection effect.

Overall, most of the lines retained in the 1mm and 0.9\,mm ranges and a large
fraction ($\simeq\,2/3$) of the lines retained in the 3\,mm and 2\,mm ranges
have a high signal to noise ratio ($\geqslant\,10$). The density of
lines with such high S/N is relatively constant in frequency and equal
to $\simeq\,4-5$/GHz.

When comparing with the line density quoted at the beginning of the
section, namely about 20 lines per GHz, obtained considering lines
with S/N$\geq$3 but no other filter, clearly the introduction of the
other criteria, unblended lines and, to a lesser degree, identified lines and  
E$_{\rm up} \leq$ 250\,K, severely underestimates the line content.

The line intensity spans more than three orders of magnitude, from 10\,mK to 
24\,K. The {\bf number of lines} showing an integrated intensity larger
than a given threshold is given in Fig. \ref{cumul}. For integrated
intensity ranging between 1 and 30\,K.km/s, the distribution roughly
follows a power law of slope -0.9. The power law breaks down in the
high and low end of the distribution. This slope is identical to the those 
observed by \citet{2001ApJS..132..281S, 2003A&A...407..589W, 
2005ApJS..156..127C} in their submillimeter surveys of Orion-KL. Similarly 
with what is observed in these surveys, this slope does not provide a good 
fit for the brightest and the weakest lines.

\subsection{Detected lines and species}\label{sec:detected}

Table \ref{tab-species} lists the species detected and identified in the survey, with the number 
of lines, the range of upper level energy of the lines and other observational quantities (see below).

In the frame of a survey analysis restricted to the four line criteria mentioned above,
69 different molecules (including ions) have been detected. They correspond to 32 distinct chemical 
species and include 37 rare isotopologues. Out of the $\sim$ 1000 lines of Table \ref{tab-species}, 
about half belong to only three species: CH$_3$HCO, HCOOCH$_3$ and CH$_3$OH.

Most of the 4000 lines detected in our survey belong in fact to the species already identified in this source (see Table 5), 
among which a few molecules show extremely rich spectra, with many weak and/or blended lines, not included in the 
present study.  Although we anticipate the presence of unidentified lines, their identification will require a careful 
analysis and even probably modeling of these spectra.  Also, with the E$_{up}$ $\leq$ 250\.K selection criterion used in this 
work, we do not identify any vibrationally excited lines. These objectives are postponed to a future article. 
%
%
\begin{table*}[!th] 
\small{
\caption{Detected species ordered by decreasing fluxes of the main isotopologue.}
\label{tab-species} 
\begin{centering} 
\begin{tabular}{l@{\hspace{0.15cm}}l@{\hspace{0.12cm}}l@{\hspace{0.12cm}}r@{$-$}l@{\hspace{0.12cm}}c@{$\pm$}c@{\hspace{0.12cm}}c@{$\pm$}l@{\hspace{0.001cm}}r|l@{\hspace{0.15cm}}l@{\hspace{0.12cm}}l@{\hspace{0.12cm}}r@{$-$}l@{\hspace{0.12cm}}c@{$\pm$}c@{\hspace{0.12cm}}c@{$\pm$}l@{\hspace{0.0001cm}}r}
\hline 
\hline 
TAG & Species & Nb$^{(2)}$  & \multicolumn{2}{c}{E$_{\rm up}$ (K)} & \multicolumn{5}{c|}{ {\hspace{-0.3cm}} $<$V$_{\rm LSR}$$>${\hspace{0.1cm}}$<$FWHM$>${\hspace{0.01cm}} $\int T_{\rm mb} dv$} &  
TAG & Species & Nb$^{(2)}$  & \multicolumn{2}{c}{E$_{\rm up}$ (K)} & \multicolumn{5}{c}{ {\hspace{-0.3cm}}  $<$V$_{\rm LSR}$$>${\hspace{0.1cm}}$<$FWHM$>${\hspace{0.01cm}} $\int T_{\rm mb} dv$} \\
$^{(1)}$     &   $^{(3)}$           & lines & \multicolumn{2}{c}{min-max} & \multicolumn{5}{c|}{ {\hspace{0.35cm}} (km.s$^{-1}$) {\hspace{0.4cm}} (K.km.s$^{-1})$ }  &
$^{(1)}$     &       $^{(3)}$       & lines & \multicolumn{2}{c}{min-max}  & \multicolumn{5}{c}{ {\hspace{0.35cm}} (km.s$^{-1}$) {\hspace{0.4cm}}  (K.km.s$^{-1})$ }   \\
\hline 
28503	&$^{(aa)}$CO 				&3		&5.5		&33.2	&5.93	&0.71	&9.47	&1.77	&605.8 	& 41001	&$^{(bb)}$CH$_3$CN                &$>$30  	&20.4	&21.2	& \multicolumn{2}{c}{ } & \multicolumn{2}{c}{ }&58.6 \\
29501	&$^{(ab)}$$^{13}$CO		&3		&5.3		&31.7	&3.94	&0.36	&4.13	&1.16	&116.0 	& 46509	&$^{(bc)}$H$_2$CS 		         &26		&9.9		&143.4	&3.61	&0.35	&4.53	&1.34	&51.2 \\
30502	&$^{(ac)}$C$^{18}$O 		&3		&5.3		&31.6	&3.72	&0.18	&2.90	&0.61	&67.6   	& 47505	&$^{(bd)}$H$_2$$^{13}$CS     &4		&22.5	&97.1	&4.09	&0.79	&4.61	&2.29	&2.9 \\
29006	&$^{(ad)}$C$^{17}$O		&3		&5.4		&32.4	&3.84	&0.35	&3.75	&0.66	&25.5	& 47504	&$^{(be)}$HDCS 			&5		&8.9		&30.9	&3.77	&0.12	&2.81	&0.88	&0.8 \\
48501	&$^{(ae)}$SO 				&20		&9.2		&87.5	&3.98	&0.20	&4.18	&0.55	&339.5	& 60503	&$^{(bf)}$OCS 			&12		&16.3	&134.8	&3.13	&0.33	&5.12	&0.77	&38.6 \\
50501	&$^{(af)}$$^{34}$SO		&13		&15.6	&86.1	&3.81	&0.27	&4.87	&0.84	&28.0	& 62505	&$^{(bg)}$OC$^{34}$S		&9		&15.9	&157.2	&2.51	&0.26	&3.98	&0.65	&7.8 \\
30591	&$^{(ag)}$o-H$_2$CO		&10		&6.8		&143.3	&3.45	&0.58	&4.09	&0.58	&163.3	& 61502	&$^{(bg)}$O$^{13}$CS		&10		&20.9	&147.2	&2.58	&0.51	&4.49	&1.25	&5.6 \\	
30581	&$^{(ag)}$p-H$_2$CO		&7		&10.5	&99.7	&3.67	&0.12	&3.92	&0.65	&93.6	& 60003	&$^{(bh)}$HCOOCH$_3$		&182	&17.9	&145.0	&2.52	&0.32	&2.71	&0.77	&51.6 \\
31501	&$^{(ah)}$HDCO 			&10		&17.6	&102.6	&3.39	&0.58	&3.99	&1.06	&26.7	& 44505	&$^{(bi)}$SiO				&6		&6.2		&75.0	&4.40	&0.46	&5.57	&0.78	&49.7 \\
31503	&$^{(ai)}$H$_2$$^{13}$CO	&12		&10.2	&157.5	&2.96	&0.38	&2.86	&1.29	&12.7  	& 27002	&$^{(bj)}$HNC 			&3		&4.3		&43.5	&3.80	&0.56	&2.81	&1.08	&32.1 \\
32592	&$^{(aj)}$o-D$_2$CO		&6		&16.8	&81.2	&3.51	&0.61	&3.80	&0.84	&11.7	& 28508	&$^{(bk)}$DNC			&2		&11.0	&22.0	&4.31	&0.02	&2.47	&1.19	&4.9 \\
32582	&$^{(aj)}$p-D$_2$CO		&8		&5.3		&99.5	&3.38	&0.72	&2.76	&1.24	&5.5	 	& 28005	&$^{(as)}$HN$^{13}$C		&4		&4.2		&41.8	&3.74	&0.61	&3.44	&1.76	&4.8 \\
32503	&$^{(ak)}$H$_2$C$^{18}$O	&4		&20.0	&63.4	&2.71	&0.16	&2.67	&1.21	&1.9		& 51501	&$^{(bl)}$HC$_3$N 			&17		&19.6	&203.0	&3.53	&0.44	&4.67	&1.97	&35.4 \\
64502	&$^{(al)}$SO$_2$ 			&82		&7.7		&248.5	&3.74	&0.32	&5.30	&1.22	&238.3 	& 52005	&$^{(bm)}$DC$_3$N		&6		&22.3	&77.0	&4.29	&0.23	&2.61	&0.41	&1.0 \\
66002	&$^{(am)}$$^{34}$SO$_2$	&5		&7.6		&69.7	&3.33	&0.72	&4.20	&1.71	&1.6		& 34502	&$^{(bn)}$H$_2$S 			&2		&27.9	&84.0	&3.55	&0.52	&4.90	&0.08	&25.2 \\
32504	&$^{(an)}$CH$_3$OH 		&99		&7.0		&233.6	&3.42	&0.42	&5.91	&1.35	&217.6	& 35001	&$^{(bo)}$HDS 			&2		&11.7	&34.7	&3.05	&0.55	&2.41	&2.09	&0.7 \\
33502	&$^{(ao)}$$^{13}$CH$_3$OH 	&9		&6.8		&94.6	&2.93	&0.47	&3.92	&1.63	&2.8 	& 38082	&$^{(bp)}$o-c-C$_3$H$_2$	&14		&4.1		&48.3	&4.10	&0.27	&2.73	&0.70	&14.5 \\           		
		&$^{(ap)}$CH$_2$DOH		&12 	       	&4.5		&40.8	&2.74	&0.64	&5.89	&0.91	&6.4 	& 38092	&$^{(bp)}$p-c-C$_3$H$_2$	&9		&6.4		&47.5	&4.27	&0.26	&2.57	&1.29	&4.5 \\
          	&$^{(ap)}$CHD$_2$OH		&7		&4.2		&53.5	&3.70	&0.88	&4.76	&1.58	&2.6 	& 39003	&$^{(bq)}$c-C$_3$HD 		&5		&17.4	&25.34	&4.44	&0.31	&1.91	&0.88	&1.5 \\ 
           	&$^{(ap)}$CH$_3$OD		&4		&6.0		&26.8	&2.55	&0.84	&3.96	&1.29	&2.3 	& 46008	&$^{(br)}$CH$_3$OCH$_3$	&82		&11.1	&169.8	&2.53	&0.29	&2.27	&0.69	&18.8 \\
          	&$^{(aq)}$CD$_3$OH		&3		&12.2	&42.0 	&2.43	&0.21	&2.03	&0.49	&0.5		& 40502	&$^{(bs)}$CH$_3$CCH 		&26		&12.3	&176.6	&3.62	&0.32	&3.03	&0.58	&17.4 \\
29507	&$^{(ar)}$HCO$^+$ 		&3		&4.3		&42.8	&3.29	&0.16	&2.87	&1.39	&144.2 	& 41502	&$^{(bt)}$CH$_2$DCCH		&4		&16.3	&38.1	&4.34	&0.31	&1.93	&0.40	&0.2 \\
30504	&$^{(ar)}$H$^{13}$CO$^+$	&4		&4.2		&41.6	&3.94	&0.46	&2.43	&0.44	&31.5 	& 25501	&$^{(bu)}$CCH 			&13		&4.2		&25.1	&3.83	&0.09	&2.08	&0.20	&14.7 \\
30510	&$^{(as)}$DCO$^+$ 		&3		&10.4	&51.9	&4.19	&0.29	&2.34	&0.83	&14.3 	& 26501	&$^{(bv)}$CCD 			&4		&10.4	&10.3	&4.32	&0.52	&2.54	&0.86	&0.8 \\
31506	&$^{(at)}$HC$^{18}$O$^+$	&3		&4.1		&24.5	&4.04	&0.37	&2.15	&0.51	&1.6 	& 43002	&$^{(bw)}$HNCO 			& $>$8 	&10.5	&143.5	&\multicolumn{2}{c}{ } & \multicolumn{2}{c}{ } &13.8 \\
31508	&$^{(ar)}$D$^{13}$CO$^+$	&2		&10.2	&20.4	&4.52	&0.16	&2.34	&1.36	&0.6 	& 26504	&$^{(bx)}$CN				&7		&5.4		&5.4		&4.45	&0.10	&1.82	&0.16	&8.5 \\
30505	&$^{(au)}$HC$^{17}$O$^+$	&1		&12.5	&12.5	&3.94	&0.22	&4.03	&0.71	&0.4		& 45506	&$^{(by)}$HCS$^+$	&5		&6.1		&73.7	&3.72	&0.20	&3.32	&0.70	&5.1 \\
44501	&$^{(av)}$CS 				&4		&7.0		&65.8	&3.70	&0.12	&3.56	&0.43	&126.5	& 42501	&$^{(bz)}$H$_2$CCO		&13		&9.7		&66.9	&3.00	&0.26	&2.58	&0.44	&4.8 \\
46501	&$^{(av)}$C$^{34}$S		&4		&6.9		&64.8	&3.63	&0.16	&3.60	&0.68	&21.1	& 18501	&$^{(ca)}$NH$_2$D			&$>$3	&16.6	&182.8	& \multicolumn{2}{c}{ } & \multicolumn{2}{c}{ } &4.7 \\
45501	&$^{(av)}$$^{13}$CS		&4		&6.7		&46.6	&3.69	&0.22	&4.06	&0.96	&9.0		&  30008	&$^{(cb)}$NO				&7		&7.2		&36.1	&4.02	&0.12	&2.25	&1.42	&3.7 \\
45502	&$^{(av)}$C$^{33}$S		&4		&7.0		&65.3	&3.68	&0.66	&5.05	&1.67	&7.2		& 56007	&$^{(cc)}$C$_2$S			&12		&15.4	&53.8	&4.10	&0.17	&2.12	&0.62	&2.6 \\
27501	&$^{(aw)}$HCN 			&3		&4.2		&42.5	&3.57	&0.78	&7.06	&2.12	&97.5	& 45013	&$^{(cd)}$PN				& $>$3 	&13.5	&63.1	& \multicolumn{2}{c}{ } & \multicolumn{2}{c}{ } &2.1 \\
28002	&$^{(ax)}$H$^{13}$CN		&5		&4.1		&41.4	&3.77	&0.26	&4.05	&1.85	&17.1 	& 19002	&$^{(ce)}$HDO			&1		&95.2	&95.2	&2.87	&0.21	&6.13	&0.58	&2.1 \\
28509	&$^{(ay)}$DCN 			&3		&10.4	&52.1	&3.63	&0.46	&4.09	&0.55	&11.3 	& 46010	&$^{(cf)}$NS				&$>$6	&9.9		&17.8	& \multicolumn{2}{c}{ } & \multicolumn{2}{c}{ } &1.3 \\
28003	&$^{(az)}$HC$^{15}$N		&4		&4.1		&41.3	&3.60	&0.33	&5.52	&1.47	&6.4		& 49503	&$^{(cg)}$C$_4$H			&6		&20.5	&30.2	&4.00	&0.08	&1.58	&0.28	&0.7 \\
44003	&$^{(ba)}$CH$_3$CHO		&192	&13.8	&194.5	&2.79	&0.39	&2.68	&1.00	&70.0	& 37003	&$^{(ch)}$c-C$_3$H 		&3		&4.39	&10.8	&4.42	&0.22	&1.79	&0.10	&0.3   \\
\hline 
\hline 
\vspace{0.01cm}
\end{tabular} 
\end{centering} 
$^{(1)}$ In the CDMS as in the JPL catalog, the species are associated to a five digit number, called TAG;  the first two digits correspond to their molecular weight in atomic mass units, and the third digit is a 5 in the CDMS catalog, and a 0 in the JPL catalog. Our identification of the ortho and para forms of H$_2$CO, D$_2$CO and c-C$_3$H$_2$ relies on the VASTEL Spectroscopic database (http://www.astro.caltech.edu/~vastel/CHIPPENDALES) so that these species are associated with specific TAG. The deuterated isotopologues of methanol, not yet included in the CDMS and JPL databases, do not have a TAG. Our identification of their lines is based on the data given in  \cite{2002A&A...393L..49P} and \cite{2004A&A...416..159P} and references therein.\\
$^{(2)}$ For most N-bearing species, due to unresolved hyperfine structure, this number is a lower limit corresponding to groups of blended lines. \\
$^{(3)}$ For each species, the spectroscopic references given here are the most recent cited in the CDMS and JPL databases. All of them should be read as reference and references therein: 
$^{aa}$\citet{1997JMoSp.184..468W}; 
$^{ab}$\citet{2004ApJ...611..615C}; 
$^{ac}$\citet{2001ZNatA..56..329K}; 
$^{ad}$\citet{2003ApJ...582..262K}; 
$^{ae}$\citet{1997JMoSp.182...85B}; 
$^{af}$\citet{1996JMoSp.180..197K}; 
$^{ag}$\citet{2000JMoSp.200..143M}; 
$^{ah}$\citet{1999JMoSp.195..345B}; 
$^{ai}$\citet{2000PCCP....2.3401M}; 
$^{aj}$\citet{2004JMoSp.228....1L};  
$^{ak}$\citet{2000ZNatA..55..486M}; 
$^{al}$\citet{2000JMoSp.201....1M}; 
$^{am}$\citet{1998JMoSp.191...17B};
$^{an}$\citet{2004A&A...428.1019M}; 
$^{ao}$\citet{1997JPCRD..26...17X};  
$^{ap}$\citet{2002A&A...393L..49P};  
$^{aq}$\citet{2004A&A...416..159P};  
$^{ar}$\citet{2007ApJ...662..771L}; 
$^{as}$\citet{2009A&A...507..347V};
$^{at}$\citet{2004A&A...419..949S};
$^{au}$\citet{2001CaJPh..79..359D};
$^{av}$\citet{2003JMoSp.219..296K};
$^{aw}$\citet{2003ApJ...585L.163T};   
$^{ax}$\citet{2005JMoSp.233..280C}; 
$^{ay}$\citet{2004JMoSp.225..152B};   
$^{az}$\citet{2005ApJS..159..181C};   
$^{ba}$\citet{1996JPCRD..25.1113K};   
$^{bb}$\citet{2009A&A...506.1487M};   
$^{bc}$\citet{1994JChPh.101.7300C};   
$^{bd}$\citet{1987MolPh..62.1429B};   
$^{be}$\citet{1997ApJ...491L..63M};   
$^{bf}$\citet{2005JMoSp.234..190G};   
$^{bg}$\citet{1987JChPh..87.2010L};   
$^{bh}$\citet{2008JMoSp.251..293M};   
$^{bi}$\citet{1998ApJ...496L..51C};   
$^{bj}$\citet{2000A&A...363L..37T};   
$^{bk}$\citet{2006JMoSt.780....3B};   
$^{bl}$\citet{2000JMoSp.204..133T};   
$^{bm}$\citet{2008CP....346..132S};
$^{bn}$\citet{1995JMoSp.173..380B};
$^{bo}$\citet{1985JMoSp.109..300C};
$^{bp}$\citet{1993JChPh..99..890M};
$^{bq}$\citet{1987JMoSp.122..313B};
$^{br}$\citet{1990Naturforsch.45a..702B};
$^{bs}$\citet{2008A&A...487.1197C};
$^{bt}$\citet{1993JMoSp.160..471L};
$^{bu}$\citet{2009A&A...505.1199P};
$^{bv}$\citet{1985A&A...144L..15B};
$^{bw}$\citet{2007AstL...33..121L};
$^{bx}$\citet{1995A&A...304L...5K};
$^{by}$\citet{2003PCCP....5.2770M};
$^{bz}$\citet{2003Naturforsch.58a..275B};
$^{ca}$\citet{1988JMoSp.127..240F};
$^{cb}$\citet{1972JMoSp..44..320M};
$^{cc}$\citet{1992ApJ...399..325L};
$^{cd}$\citet{2006JMoSt.780..260C};
$^{ce}$\citet{1985JOSAB...2.1340J};
$^{cf}$\citet{1982JMoSp..93..416L};
$^{cg}$\citet{1995JChPh.103.7828C};
$^{ch}$\citet{1994JChPh.101.5484Y};}
\end{table*} 

Table \ref{tab-species} also lists the sum of the line flux in
each species. For species showing rich and complex spectra or with
numerous lines with upper energy levels $\geq$ 250\,K, this value is in
fact a lower limit due to the numerous weak and blended lines not
included in the present analysis. 


As noted by the previous study
\citep{1994ApJ...428..680B, 1995ApJ...447..760V}, the millimeter and
submillimeter spectrum of IRAS16293 is dominated by simple O-rich
species like CO, SO, H$_2$CO, SO$_2$, and CH$_3$OH (the three first
families alone contribute to more than 2/3 of the total flux). In the
frequency range covered by our survey, the total flux emitted in the
CO main and isotopic lines is about 800 K.km.s$^{-1}$, $\sim$ 30\%
smaller than the flux emitted in the SO, H$_2$CO, SO$_2$, and CH$_3$OH
lines together ($\sim1100$\,K.km.s$^{-1}$). Thus CO is not the major
cooling agent in this frequency range.  In addition, our survey shows
the presence of thousands of weaker lines from heavier and more complex
molecules not detected in the previous surveys. The blow-ups of
Fig. \ref{survey} illustrate the situation, with several lines from
CH$_3$CCH, CH$_3$CN, HCOOCH$_3$ etc. To have a rough estimate of the
contribution of this ``grass'' of lines to the total cooling of the
gas (in this range of frequencies), consider a line density of about
15 lines per GHz (see above) and a line integrated intensity of $\sim
0.3$\,K.km.s$^{-1}$. This gives approximatively an integrated line
intensity of 900\,K.km.s$^{-1}$, namely comparable to the contribution of CO and its
isotopologues.

\subsection{Line parameters}\label{sec:line-kinematics}

A Gaussian fit has been performed for each of the lines of Table
\ref{lines-on-line}.
The parameters of the fit (integrated intensity,
line width FWHM and rest velocity v$_{\rm LSR}$) are reported in the same
table. The uncertainties on the integrated line intensity take into
account the spectra noise and the calibration uncertainties
reported in Table \ref{tab-observations}. Note that we have verified
that even when the line profile differs from a simple Gaussian (for
example in self-absorbed lines, or lines with broad wings) the Gaussian fit area 
is very close to the true integrated line intensity and, therefore, reliable. The
uncertainties on the line width FWHM and rest velocity v$_{\rm LSR}$ take
into account the statistical errors (from the fit) and the uncertainty
due to the limited spectral resolution, which indeed dominates the
error. The cases when the line profiles are clearly not Gaussian are
marked in Table \ref{lines-on-line}. 
The average $<$FWHM$>$
and $<$v$_{\rm LSR}>$ for each species and isotopologues are reported in Table \ref{tab-species}, except for those which
show severe blending due to unresolved or partially resolved hyperfine structure (CH$_3$CN, HNCO, NH$_2$D, NS).

In the following, we analyze the information derived from the Gaussian FWHM and v$_{\rm LSR}$ of the lines. 

%
%

\begin{figure*}[htb]
\centering
\includegraphics[width=0.95\linewidth]{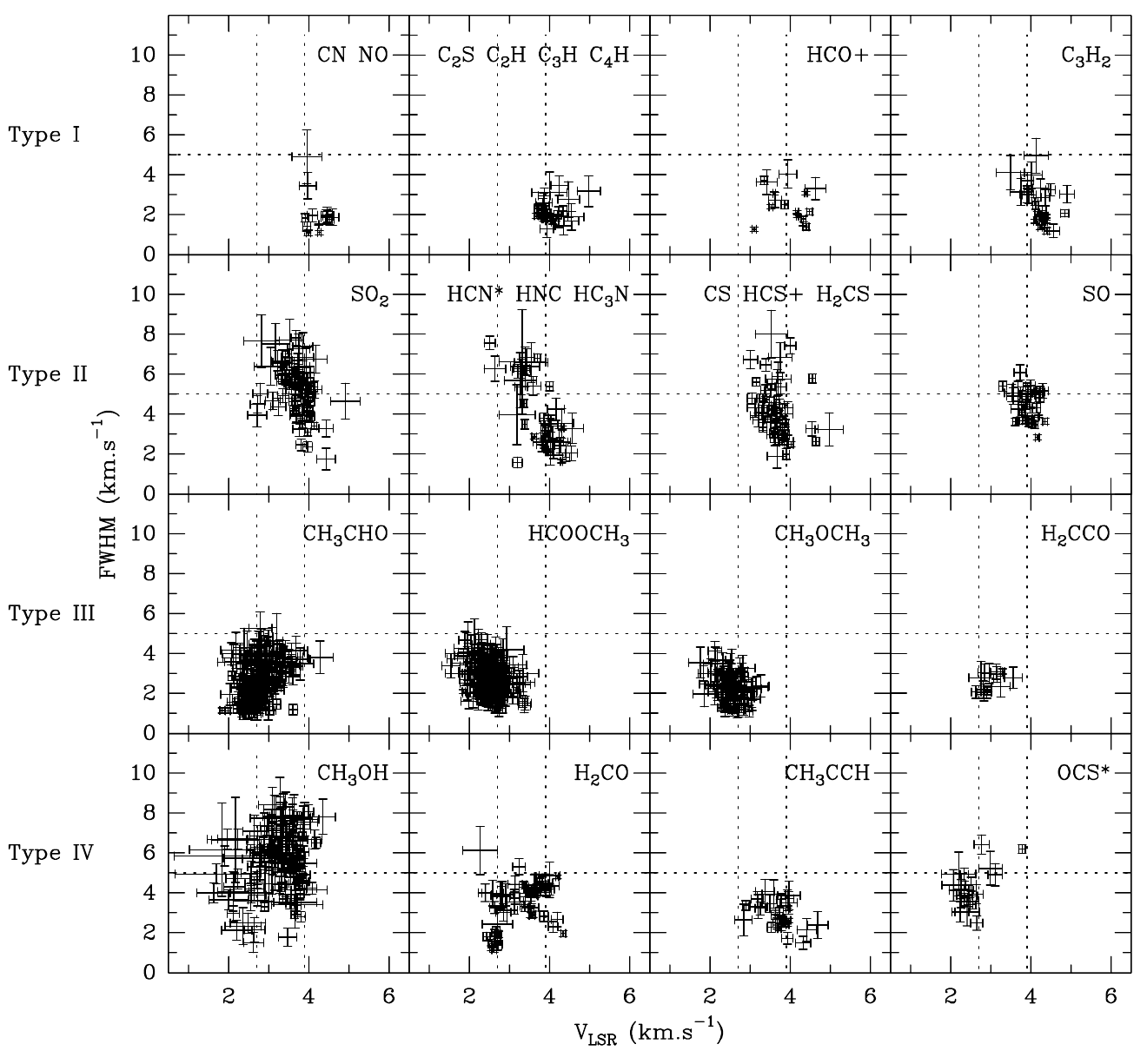}
\caption{Plots of the rest velocity v$_{\rm LSR}$ versus the FWHM, derived from the Gaussian fits 
  of the lines (Table \ref{lines-on-line}). All the detected species and the relevant isotopologues of 
  Table \ref{tab-species} are plotted, except those in which the lines have obviously non Gaussian profiles 
  (see text). In particular, the labels HCN* and OCS* mean that the main isotopologues of these species are 
  not included, due to their non Gaussian profiles. The 1 sigma errorbars include fit and spectral resolution 
  uncertainties. The vertical lines at v$_{\rm LSR}$\,=\,2.7 and 3.9\,km.s$^{-1}$ correspond to the velocity 
  of the components B and A respectively (see \S \ref{sec:discussion}). The horizontal line at 5\,km.s$^{-1}$ 
  corresponds approximatively to the average of the line FWHM range.}
\label{fig-FW-Vo}
\end{figure*}

Figure \ref{fig-FW-Vo} shows the FWHM versus v$_{\rm LSR}$
for each species. In the plots, we have regrouped the isotopologues of
the same species and, in a few cases, species with a small number of
lines. Note that in these cases we verified that the species have
similar FWHM and v$_{\rm LSR}$ to avoid introducing artificial trends in
the plot. Figure \ref{fig-FW-Vo} shows a remarkable and unexpected
behavior: {\it the FWHM and v$_{\rm LSR}$ distributions are not the same
  but, on the contrary, they depend on the species}. 

Based on the
different FWHM and v$_{\rm LSR}$ distributions, it is possible to identify
four types of "kinematical behaviors":
\begin{enumerate}
\item Type\,I (first row of Fig. \ref{fig-FW-Vo}): $<$v$_{\rm LSR}$$>$\,$\sim$\,4.0\,km.s$^{-1}$ 
 and $<$FWHM$>$\,$\sim$\,2.5\,km.s$^{-1}$. The lines
  show very little dispersion both in terms of rest velocities and in
  term of line widths.  Small carbon chains and rings, and "small
  molecules" belong to this group.
\item Type\,II (second row): $<$v$_{\rm LSR}$$>$\,$\sim$\,3.7\,km.s$^{-1}$
  with a very little dispersion, FWHM from $\sim$2\,km.s$^{-1}$ to
  $\sim$8\,km.s$^{-1}$. All species in this group are S- or N- bearing
  molecules. It can be noticed that HCO$^+$ and C$_3$H$_2$ (in the
  first row) show a behavior intermediate between Type\,I and Type\,II.
\item Type\,III (third row): $<$v$_{\rm LSR}$$> $$\sim$ 2.5--3.0\,kms$^{-1}$,
  $<$FWHM$>$ $\leq$ 4.0\,km.s$^{-1}$. Four complex organic molecules
  show this behavior.
\item Type\,IV (fourth row): v$_{\rm LSR}$ and FWHM showing a mixed
  behavior, with characteristics belonging to the two last
  groups. CH$_3$OH lines have v$_{\rm LSR}$ and FWHM ranging from 2 to
  9\,km.s$^{-1}$; H$_2$CO and CH$_3$CCH lines have moderate FWHM 
  (4--5\,km.s$^{-1}$) and v$_{\rm LSR}$ ranging from 2.5 to 4\,km.s$^{-1}$; lines from
  the rare isotopes of OCS have v$_{\rm LSR}$\,$\sim$\,2.5\,km.s$^{-1}$ and 
  FWHM\,$\leq$\,4\,km.s$^{-1}$.
\end{enumerate}

It can be noted that the FWHM averaged over each of
the four frequency bands of the survey is similar, between 2 and 5\,kms$^{-1}$, with a slight increase in 
the 1\,mm band due to a poorer spectral resolution.
We have verified that none of the four kinematical behaviors is an artifact due to this instrumental effect.
Table \ref{tab:types} summarizes the situation.  
In order to better understand the physical meaning of the plots in
Fig. \ref{fig-FW-Vo} (and of the four identified Types) we have plotted  in Fig. \ref {fig-FW-Eup}
the values of the FWHM as function of the upper level energy 
E$_{\rm up}$ of the transition. 
The species have been grouped as in Fig. \ref{fig-FW-Vo}; it is striking that the distinction 
between the four Types defined by the (FWHM, v$_{\rm LSR}$) distribution is visible also in this plot.
%
%
%
\begin{table}[!th]
\caption{Distribution of the detected species in four kinematical types.}
\label{tab:types}
\begin{center}
\begin{tabular}{ccc@{   }ccc}
\hline
\hline\\[-1.8ex]
 Type &  v$_{\rm LSR}$ & FWHM & E$_{\rm up}$ & FWHM  & Species \\
(Source)  & \multicolumn{2}{c}{(kms$^{-1}$)} &(K)   & behavior    & \\[0.5ex]
\hline\\[-1.8ex]
       & &                       &                    &                                         &CN, NO, \\
Type\,I & $\sim$ 4       & $\leq 2.5$ &  0--50 & constant            & C$_2$S, C$_2$H, \\
(Envelope) & &                       &                    &                    &  C$_3$H,  C$_4$H,\\         
      & &                       &                    &           &         HCO$^+$, C$_3$H$_2$\\[0.5ex]
\hline\\[-1.8ex]
       &  &                       &                    &                                         & HCN, HC$_3$N, \\
Type\,II & $\sim$\,3.7   & 2--8   & 0--250 & increases       &  HNC,  \\
(Source A) & &                       &                    &                            &  SO, SO$_2$, CS,   \\
      & &                       &                    &                        &             HCS$^+$,  H$_2$CS \\[0.5ex]
\hline\\[-1.8ex]
       & &                       &                    &                                        & CH$_3$CHO,\\
Type\,III & 2.5--3     & $\leq 4$    & 0--200       &             constant          & HCOOCH$_3$, \\
(Source B)  & &                       &                    &                       &CH$_3$OCH$_3$, \\ 
      & &                       &                    &                       & H$_2$CCO \\[0.5ex]
\hline\\[-1.8ex]
    &    &                       &                    &                            & CH$_3$OH,  \\  
Type\,IV &   2.5--4     & 2--8                 & 0--250 &    increases             & H$_2$CO,  \\
(mixed) & &                       &                    &                                          & CH$_3$CCH,  \\
      & &                       &                    &                       & OCS \\ [0.5ex]

\hline \hline
\end{tabular}
\end{center}
\noindent First column reports the associated component (see \S \ref{sec:discussion}), 
second, third and fourth columns report the typical velocities, FWHM and upper level 
energy E$_{\rm up}$ ranges of the detected lines. Fifth column describes the behavior 
of FWHM with increasing E$_{\rm up}$ and last column lists the species belonging to
 each type.
 \end{table}

Type\,I species have lines with E$_{\rm up}$ lower than 50\,K. It should be noted that this is not an observational bias:
excepted in a few cases of very light molecules, such as C$_2$H, the species associated to Type\,I present high 
energy transitions in the survey frequency range, but the line intensities decrease abruptly when E$_{\rm up}$ becomes larger than 50\,K.
In contrast, Type\,II, III and IV species present lines with E$_{\rm up}$ up to 200\,K but show different behaviors. For Type\,II species, the FWHM
increases with E$_{\rm up}$, whereas for Type\,III and IV the FWHM is constant and does not depend on E$_{\rm up}$. 
In contrast, analogous plots of v$_{\rm LSR}$ vs E$_{\rm up}$ show that the lines' velocity does not depend on E$_{\rm up}$ in any species. 
A related effect has already been observed by \cite{2002A&A...390.1001S}, who noted a correlation between the 
linewidths and the excitation temperatures derived by \cite{1994ApJ...428..680B} and \cite{1995ApJ...447..760V}. 
These properties will be used to give a physical meaning to the four Types in \S \ref{sec:discussion}.  Finally, it can be noticed that 
when detected, the deuterated species show the same behaviour as the main isotopomers, except the D-isotopomers of Type III 
species that anyway present too weak lines to be detected in our survey.

%

\section{Discussion}\label{sec:discussion}

\subsection{Comparison with previous surveys}

When compared to the previous survey toward IRAS16293
\citep{1994ApJ...428..680B, 1995ApJ...447..760V}, the present one not
only enlarges the covered frequency range ($\sim$\,200\,GHz versus $\sim$\,40\,GHz) 
but also the number of detected species, thanks to the higher sensitivity 
($\sim$\,10\,mK versus $\sim$\,40\,mK). The average line density of the 
previous unbiased survey of IRAS16293 was 7 lines per GHz to compare 
with 20 lines per GHz for the present one. Several species detected in our survey were
not detected in the previous one: complex organic molecules
(HCOOCH$_3$, HCOOCH$_3$ and CH$_3$OCH$_3$), carbon chains and rings
(C$_2$S, C$_ 4$H, c-C$_3$H), N-bearing species (NO, PN, NS) and
several D-bearing molecules.

Towards hot cores, numerous surveys have been performed. They cover
the whole range of frequencies reachable from the ground, from the 3\,mm
range observed with IRAM, SEST, NRAO or JCMT to the submillimeter
windows observable with the CSO and JCMT \citep{1996A&AS..119..333M,
    1997ApJS..108..301S, 1997A&AS..124..205H, 1998ApJS..117..427N,
    1999A&AS..135..531T, 2000ApJS..131..483K, 2000ApJS..128..213N,
    2001ApJ...551..333L, 2001ApJS..132..281S, 2003A&A...407..589W,
    2005ApJS..156..127C, 2006ApJS..162..161K, 2007msl..confE..10B,
    2007A&A...476..791O, 2010A&A...517A..96T}. 
  The line densities usually range from 10 to 20 lines per GHz,
  i.e. comparable to the present survey. SgrB2 appears as a noticeable
  exception, with significantly higher line densities, as high as 100
  lines per GHz in the 3\,mm range observed with IRAM
  \citep{2007msl..confE..10B} or 40 lines per GHz in the 1\,mm range
  observed with SEST 
  \citep{1998ApJS..117..427N,2000ApJS..128..213N}. Interestingly, the
  slope of -0.9 that we observe for the cumulative number of lines
  versus flux threshold is identical to the slopes observed by
  \citet{2001ApJS..132..281S, 2003A&A...407..589W,
    2005ApJS..156..127C} in their submillimeter surveys of Orion-KL.

In conclusion, in terms of molecular content, our survey reveals a
richness comparable to that of hot cores and confirms that the high
abundance of deuterated isotopologues, which are easily detected for a
number of species, is a distinctive characteristics of low mass protostars.

\subsection{Kinematical types and associated components}
As mentioned in the Introduction, IRAS16293 is formed by a
proto-binary system surrounded by an infalling envelope. In addition,
multiple outflows originate from the system (e.g. \citealt{2001A&A...375...40C}). 
Before attempting to interpret the observations of \S \ref{sec:line-kinematics} and, 
specifically, the meaning of the four kinematical types of Table \ref{tab:types},
based on the line rest velocities and widths, we review what is known
so far about the envelope and the proto-binary system.\\

\noindent
{\it a) Envelope}\\[0.9ex]
The envelope extends for 6000--7000\,AU in radius, equivalent to about
100$"$ in diameter, and it is relatively massive ($\sim 2$\,M$_\odot$)
\citep{2010A&A...519A..65C}. At the border of the envelope the dust
temperature is $\sim$\,13\,K and the density is $\sim 10^5$\,cm$^{-3}$. 
The dust temperature reaches 100\,K at a radius 75--86\,AU,
where the density is (2--3)$\times\,10^8$\,cm$^{-3}$, creating the region
called hot corino. The rest velocity of the cold envelope has been
measured in several studies and it is $\sim$3.9\,km.s$^{-1}$ \citep{1990ApJ...356..184M,
2004ApJ...617L..69B, 2007ApJ...662..431T}. Molecules probing the
cold envelope have $\sim 2$\,km.s$^{-1}$ line widths.\\

%
%
%
\begin{table}[htb]
\caption{Correspondence between kinematical types and spatial distributions derived from interferometric observations }
\label{tab-type-comp} 
\begin{center}
\begin{tabular}{clllc}
\hline \hline\\[-1.8ex]
Type	 & Source A  & Source B  & Envelope  & Refs.	\\[0.5ex]
\hline\\[-1.8ex]
Type\,I      &  	           & (C$_3$H$_2$) & CN          & B2004	  \\[0.5ex]
\hline\\[-1.8ex]
              &  HCN        & (HCN)       &                & T2007 	  \\
                & HCN         &                 &                & H2005	  \\
                & HC$_3$N  &                 &                & C2005	 \\
                & SO            & (SO)	       &	          & C2005	 \\
Type\,II    & SO 	    &		       &	          & H2005	 \\
		& SO$_2$	    &		       &                 & C2005\\
                & SO$_2$     & (SO$_2$)	&     	& H2005	 \\
                & H$_2$CS   & (H$_2$CS)	&     	& H2005	  \\
		& 		    & 		&	      & K2004	 \\[0.5ex]
\hline\\[-1.8ex]
               & 		    & CH$_3$CHO&	        & B2008  \\
	      &			  & CH$_3$OCH$_3$&	& C2005  \\
                & (CH$_3$OCH$_3$) & CH$_3$OCH$_3$	&	& H2005  \\
                &				& H$_2$CCO		&	& K2004  \\
Type\,III  & (H$_2$CCO)		& H$_2$CCO		&	& B2008  \\
                & 		    & HCOOCH$_3$&	& R2006  \\
                & HCOOCH$_3$ & (HCOOCH$_3$)	&	& B2004	\\
	        & HCOOCH$_3$ & HCOOCH$_3$	&	& K2004	\\
                & HCOOCH$_3$ &				&	& C2005	 \\[0.5ex]
\hline\\[-1.8ex]
               &  CH$_3$OH	&			&	& C2005	\\
               & CH$_3$OH	& CH$_3$OH	&	& K2004	 \\
Type\,IV   & H$_2$CO	& H$_2$CO	&	& S2004	 \\
               & H$_2$CO	& H$_2$CO	&	& C2005	 \\
               & OCS		& OCS		&	& H2005	 \\[0.5ex]
\hline
\hline
\end{tabular}
\end{center}
The first column reports the kinematical type of the species according to the definition given in Table \ref{tab:types}. 
The columns 2, 3 and 4 report where the species has been detected: source A, B and envelope respectively. 
The species in parenthesis means that weaker emission has also been observed in the relevant component 
or that they are questionable identifications (see text). Last column reports the interferometric observations references; B2004: 
\citet{2004ApJ...617L..69B}; B2008: \citet{2008A&A...488..959B}; C2005: \citet{2005ApJ...632..371C}; H2005: 
\citet{2005AdSpR..36..146H}; K2004: \citet{2004ApJ...616L..27K}; R2006: \citet{2006ApJ...640..842R}; S2004: 
\citet{2004A&A...418..185S}; T2007: \citet{2007ApJ...662..431T}.
\end{table}

\noindent
{\it b) Proto-binary system}\\[0.9ex]
Several interferometric studies have been carried out in the past to better
characterize the nature of the two sources, A (South-East) and B (North-West),
forming the binary system (\citet{2004ApJ...616L..27K, 2004ApJ...617L..69B, 
2005ApJ...632..371C, 2008A&A...488..959B}; see also references in Table \ref
{tab-type-comp}). 
Some species are associated only or predominantly with one of the two sources A 
and B, and others are observed in both sources. All studies agree in measuring broader 
lines towards A ($\sim\,8$\,km.s$^{-1}$) than towards B ($\sim\,3$\,km.s$^{-1}$). 
However, the situation about the rest velocity of the molecular emission in the two objects
is more confused: there is some evidence that the two objects have different rest velocities 
(higher in source A than in source B), but this difference might also be due to absorption by 
the envelope according to \citet{2005ApJ...632..371C} or  self-absorption in optically thick 
lines from source B according to \citet{2004ApJ...617L..69B}; \citet{2004ApJ...616L..27K} 
also report velocities somewhat higher in source A than in source B, with a dispersion larger 
than 2.5\,km.s$^{-1}$ from one species to another. According to  \citet{2008A&A...488..959B}, 
both A and B sources show emission at velocities between 1.5 and 2.5\,km.s$^{-1}$; 
\citet{2005AdSpR..36..146H} mention two velocity components at 1.5 and 4.5\,km.s$^{-1}$ 
for source A and show emission from B peaking at $\sim\,2$\,km.s$^{-1}$. It can be noted that all 
these studies use moderate spectral resolutions ($\sim$\,1 km.s$^{-1}$), deal with a small number 
of lines (often only one) for each species and, in some cases, suffer from poor signal to noise ratio 
for the weakest lines. 

Table \ref{tab-type-comp} summarizes,  for each of the species classified according to their 
kinematical Type in Table \ref{tab:types}, the results of interferometric observations towards the three 
components of IRAS16293 (sources A and B and envelope). \\
\begin{figure*}[htb]
\centering
\includegraphics[width=0.95\linewidth]{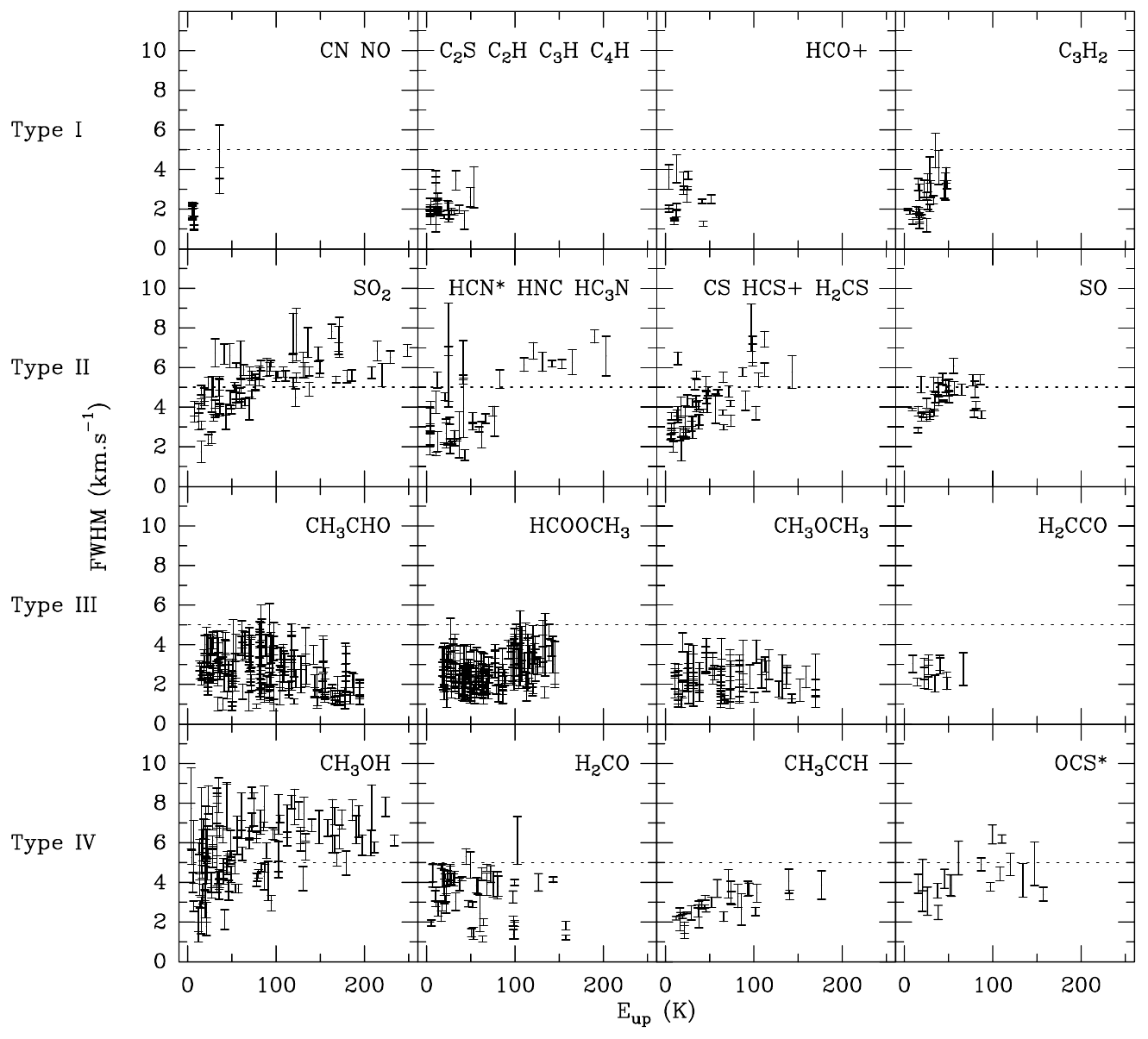}
\caption{Plots of the linewidth, FWHM, versus the energy of the upper level of the transition, E$_{\rm up}$. 
  The species are grouped as in Fig. \ref{fig-FW-Vo}. All the detected species and the relevant isotopologues of 
  Table \ref{tab-species} are plotted, except those in which the lines have obviously non Gaussian profiles (see text).
  In particular, the labels HCN* and OCS* mean that the main isotopologues of these species are not included, due 
  to their non Gaussian profiles. The 1 sigma errorbars include fit and spectral resolution uncertainties. The horizontal 
  line at 5\,km.s$^{-1}$ corresponds approximatively to the average of the line FWHM range.}
\label{fig-FW-Eup}
\end{figure*}
\noindent

{\it c) Interpretation of the kinematical types}\\[0.9ex]
Based on published interferometric observations each species is assigned to one or two of the
three components (source A, B and envelope). For example, CN has the Type\,I kinematical 
characteristics (v$_{\rm LSR} \sim$\,4\,km.s$^{-1}$ and
FWHM$\sim$\,2\,km.s$^{-1}$) and has only been detected associated with the
envelope and was not detected in any of the two sources A and B. On
the contrary, CH$_3$CHO has the characteristics of the Type\,III
behavior (v$_{\rm LSR}$\,=\,2.5--3\,km.s$^{-1}$ and FWHM\,$\leq$\,4\,km.s$^{-1}$) and has
only been detected in the direction of source B. Some species of
Table \ref{tab:types} (C$_2$H, C$_3$H, C$_4$H, C$_2$S,
NO, HNC, HCS$^+$ and CS) have not been observed with interferometers, 
to the best of our knowledge, so they are not reported in Table \ref{tab-type-comp}.
The correspondence between the four Types of Table \ref{tab:types} and the interferometric  
observations presented in Table \ref{tab-type-comp} suggests that the species belonging to the same 
kinematical Type are associated with a spatially different source: envelope (Type I), source A (Type II), 
source B (Type III) and a mix of the three previous components (Type IV). It should be noted that the 
distribution of molecules in four kinematical types is not an artifact or a bias due to the survey pointing, 
which favors emission from source B compared to emission from source A at higher frequencies. Even 
excluding all lines observed at IRAM 30\,m or JCMT telescopes with a HPBW larger than 14", i.e. all 
lines with frequencies higher than\,200 GHz, the v$_{\rm LSR}$ versus FWHM plots and the FWHM 
versus E$_{\rm up}$ plots show the same behavior.




\begin{enumerate}
\item Type\,I corresponds to molecules abundant in the cold
  envelope of IRAS16293. 
We cannot exclude that these species also emit in the densest parts of sources A and B but that 
this emission is strongly absorbed by the gas in the envelope. NO, which shows two broad and 
relatively bright lines with E$_{\rm up}$\,=\,36\,K, bright narrow lines with E$_{\rm up}$\.=\,7\,K 
and barely detected lines with E$_{\rm up}$\,=\,19\,K (not included in this paper), might be an example 
of such a situation, which deserves a detailed excitation study to come. The 343.8\,GHz line associated with source B was
  identified by \citet{2004ApJ...616L..27K} as the C$_3$H$_2$ 23$_{13, 10}$\,-\,23$_{12, 11}$ 
  transition(E$_{\rm low}$\,=\,790\,K) but the CH$_3$CHO 18$_{2, 17}$\,-\,17$_{2, 16}$ transition 
  (with E$_{\rm up}$\,=\,166\,K) seems a more reasonable identification. 
  
\item Type\,II identifies molecules prevalently associated with source A. A few exceptions, for 
which emission from source B is mentioned in the literature, deserve some attention. The identification by 
  \citet{2005AdSpR..36..146H} of a line to SO$_2$(v$_2$\,=\,1) associated with source B is also 
  questionable given the very high upper energy level (E$_{\rm low}$\,=\,1800\,K) of the transition, 
  as is the attribution by \citet{2005AdSpR..36..146H} of H$_2$CS emission to source B, as no map 
  is shown for this species. Finally, SO emission is rather extended and associated with both sources 
  A and B \citep{2005ApJ...632..371C}.  It is possible that part of the lines broadening is due to wings 
  associated with the outflow, whereas the bulk of the emission is associated with the cold envelope 
  (which would make SO rather a species of Type\,I). In addition, it should be noted that, 
without detailed excitation models which will be presented in subsequent studies,
it is not  possible to exclude that, for some Type\,II species, the cold envelope contributes to the emission 
observed at low E$_{\rm up}$ and low FWHM. 
 
 \item Type\,III species have lines prevalently emitted by source B. 
HCOOCH$_3$, which has been observed both in source A and source B appears as a notable exception. 
Excitation modeling of this species will be discussed in detail in a subsequent paper. Several qualitative arguments 
may already contribute to explain why no lines with Type II characteristics are present in this study: (i) as the lines 
emitted by the source A are much broader than the lines emitted by the source B, those lines present a higher 
percentage of blending with nearby lines which excludes them from this study; (ii) in our large beam observations, 
emission from sources A and B are superposed; the narrow lines emitted from the source B appear "on the top of" 
broad lines emitted by the source A and are, thus, more easily detected; (iii) this second effect is enhanced by the 
pointing towards the source B.
  
\item Type\,IV species are a mixed bag. The species are present in
  both source A and B and, sometimes, even in the envelope. Depending
  on the intensity contribution, the lines can have a low or high
  v$_{\rm LSR}$ and FWHM. Therefore, Type\,IV species are not
  associated with any specific component.
\end{enumerate}

Using the kinematic properties of the lines, we were able to propose general trends as whether the emission 
of the species originates from source A or source B although both sources are included in our single dish observations.
Despite the fact that the kinematic differences are small, for example a difference in the v$_{\rm LSR}$  of not more 
than 1.5\,km.s$^{-1}$, we were able to draw a general picture thanks to the large number of detected lines. 

\subsection{The dynamics of sources A and B}

Having identified where the emission from the various species
originates helps to clarify the nature of sources A and B.  As
noted in several previous works, A and B have apparently a different
chemical composition: source B is brighter in complex organic
molecules, while source A is brighter in simpler S- and N- bearing
molecules (see Table \ref{tab-type-comp}). Furthermore, it was also
noted that the FWHM of the lines arising in source A are broader than
in source B.  

Our analysis establishes two additional properties previously unrecognized (see discussion in 6.1): i) the two
sources have different velocities (v$_{\rm LSR}$) and ii) in the source A, the FWHM of the lines increases with 
increasing upper level energy (Fig. \ref{fig-FW-Eup}) whereas it remains constant in the source B. 

Concerning the linewidths behavior, a related effect, the increase of FWHM with the excitation temperature, 
had already been observed by \cite{2002A&A...390.1001S}, but this represents a less direct probe of the gas kinematics, 
as it relied on modeling assumptions to derive the excitation temperatures.
The increase of the lines FWHM  with increasing excitation has been interpreted in two ways, either this is due to the 
effect of the infalling gas on the accreting protostellar object (\cite{2003A&A...410..587C}), or the increased width is 
due to shocks caused by jet/wind interaction with the inner dense envelope (\cite{2002A&A...390.1001S}, 
\cite{2002A&A...389..908J}). 
The two interpretations predict different molecular emission distributions. However, the existing maps do not allow 
to distinguish these interpretations. In addition, to fully disentangle these two interpretations requires a detailed modeling 
of the source that is beyond the scope of this paper. At this stage, it can simply be said that the interpretation of the infalling 
gas \citep{2000A&A...357L...9C} would lead to reasonable estimates of source A and source B central masses: for optically 
thin emission, the free-fall velocity may be estimated from the linewidth, assuming that the FWHM includes quadratically a 
turbulent width $\delta _{th}$ and the free-fall broadening; its analytic expression yields to the following expression of the 
core masse, where the widths are expressed in km.s$^{-1}$ and the radius $r$ in AU:
\begin{equation}
M ({\rm M}_\odot) = 1.4\times 10^{-4} (FWHM^2  - \delta _{th}^2) r
\end{equation}
Assuming that the turbulent width in the A and B cores external layers is close to that of the envelope static gas, 
we derive $\delta _{th}$ $\sim$\,2\,km.s$^{-1}$ from the Type\,I species linewidths. According to the interferometric 
observations of Type\,II and Type\,III species reported in Table \ref{tab-type-comp}, both sources show a radius of $\sim1.5"$ 
 (i.e 180\,AU at a distance of 120\,pc). For core A, typical FHWM larger than $\sim$\,6\,km.s$^{-1}$ lead to a central 
mass of at least 0.8\,M$_\odot$, whereas for core B, where FHWM are smaller than $\sim$\,3\,km.s$^{-1}$, the 
central mass cannot exceed 0.1\,M$_\odot$. This assumes that the linewidths are due to collapsing envelopes rather than 
rotating disks. If they were, on the contrary, due to rotating disks, one would have to take into account the (unknown) inclination angle. 
The available interferometer observations, which barely spatially resolve the core A only  \citep{2004ApJ...617L..69B}, do not allow 
to distinguish between these two possibilities. In the following, we will restrict the discussion to the infalling envelopes, for 
simplicity, but our conclusions substantially apply also to the disk case.

The difference between the v$_{\rm LSR}$ of both sources can  be interpreted in the following way: the source B rotates 
around its companion, the source A; for a distance of 480\,AU between cores A and B, and a core A mass of 0.8\,M$_\odot$, 
the rotation velocity of core B is 1.2\,km.s$^{-1}$. The observed difference between A and B velocities (v$_{\rm LSR}$ of 
3.9 and 2.7\,km.s$^{-1}$ respectively) is thus perfectly consistent with this picture. \citet{2004ApJ...617L..69B} already 
discussed this possibility but did not have enough evidence for preferring it with respect to another hypothesis where the 
lines kinematics properties reflect opacity effects. The large number of lines and species observed in our survey allows to 
favor the dynamical issue.

%
%
%
\section{Conclusions}
We have presented the unbiased spectral survey of the 3, 2, 1 and 0.9\,mm
bands accessible from ground towards the Class 0 source IRAS16293.
For that, we used the IRAM-30\,m and JCMT-15\,m telescopes, during about
300 hours of observations. This is the most sensitive survey ever
published in these bands towards a solar type protostar.

The data have been released for public use in two CLASS files, which can be retrieved 
on the web site {\it http://www-laog.obs.ujf-grenoble.fr/heberges/timasss/}. The site
also contains two accompanying files (reported in the On Line
Material) providing informations about the calibration of the single
receiver settings, obtained by comparing the survey lines with
previously obtained observations.

The line density, $\sim$ 20\,lines/GHz, appears as high as in
comparable surveys obtained towards high mass protostars (with the
exception of SgrB2). More than one thousand unblended lines with
S/N$\geq 3$ and upper energy levels lower than 250\,K have been
identified through the comparison with the JPL and CDMS catalogs.
They correspond to 32 chemically distinct species, showing a chemical
richness comparable to that of hot cores. The identification of 37 additional
rare isotopologues and, specifically, numerous D-bearing molecules
confirm that IRAS16293 has a remarkably high abundance of deuterated
species. The 3\,mm\,--\,0.9\,mm spectra are dominated by relatively simple
molecules (CO, SO, H$_2$CO, SO$_2$ and CH$_3$OH). However, the
numerous weaker lines emitted by larger molecules account for at least
as much as the CO integrated line intensities.

The analysis of the profiles of this large set of identified lines, and
specifically the central velocities and widths, gives clues to
disentangle where each emission dominantly originates from: cold envelope (narrow
lines at V$_{\rm LSR} \simeq$\,3.9\,km.s$^{-1}$), source A (broader lines at
V$_{\rm LSR} \simeq$\,3.9\,km.s$^{-1}$) and source B (narrow lines at V$_{\rm LSR}
\simeq$ 2.7\,km.s$^{-1}$. Furthermore, in source A, the line widths increase
with the upper energy level of the transition.
If this behavior is interpreted as due to gas infalling towards a central object, the core A 
mass is $\sim\,1$\,M$_\odot$ and the lower line widths observed towards source B set an 
upper limit to the mass of this source, $\leq\,0.1$\,M$_\odot$. 
The observed difference in the V$_{\rm LSR}$, $\sim1.2$\,km.s$^{-1}$, is consistent with 
the source B rotating around the more massive source A. From a chemical point of view, the 
source B shows predominant emission from O-bearing complex molecules whereas N- and S- 
bearing molecules are strong emitters in the source A. In order to derive reliable estimates of 
the corresponding chemical abundances, it is necessary to carry out careful radiative transfer 
modeling. This is postponed to future articles.
%
%
\begin{acknowledgements}
  We are deeply thankful to the IRAM staff and the successive TACs,
  and particularly to Clemens Thum, for his help in preparing and
  programming the observations at the IRAM-30\,m telescope. We
  gracefully thank the JCMT staff, particularly Remo Tilanus and Jim
  Hogh, who were always able to quickly resolve problems. We
  thank Laurent Pagani for fruitful discussions about calibration
  problems. 
  We are very thankful to the molecular databases JPL and CDMS, which
  were largely used for the work presented here as to the spectroscopic groups providing the data. 
  This work has been supported by l'Agence Nationale pour la Recherche (ANR), France
  (contracts ANR-08-BLAN-022) and by the Minist\`ere de la Recherche Scientifique et Universit\'e 
  J. Fourier de Grenoble, France (PPF  WAGOS). Finally, we warmly thanks the referee, Dr, J. 
  Cernicharo, and the editor, Dr. M. Walmsley, who contributed to improve a lot this paper 
  with numerous helpful comments.
  \end{acknowledgements}
  
\bibliographystyle{aa}

\bibliography{biblio_EC}

\Online
\onllongtab{2}{
\small{
\begin{longtable}{cllrrrc}
\caption[]{Comparison of the spectral survey lines with previous observations.}
\label{tab_external_calib}\\
\hline \hline
Setting & Species & Transition & Freq. & \multicolumn{2}{c}{$\int$  T$_A^*$ $d$v} & Ref. \\
number& & &(MHz) & \multicolumn{2}{c}{(K.kms$^{-1}$)} & \\
 & & & & survey & prev. obs. & \\
\hline
\endfirsthead
\caption{Continued.} \\
 \hline
Setting & Species & Transition & Freq. & \multicolumn{2}{c}{$\int$  T$_A^*$ $d$v} & Ref. \\
number & & &(MHz) & \multicolumn{2}{c}{(K.kms$^{-1}$)} & \\
 & & & & survey & prev. obs. & \\
\hline 
\endhead
\hline
\endfoot
\hline
\endlastfoot
2       & HDO                             & 1$_{10}-1_{11}$               & 80578   & 0.507   & 0.4      & $^{(6)}$ \\
11*   & CH$_{3}$CCH           & 5$_{2}-4_{2}$                    & 85450   & 0.188   & 0.18    & $^{(3)}$ \\
11*   & CH$_{3}$CCH           & 5$_{1}-4_{1}$                    & 85455   & 0.411   & 0.5      & $^{(3)}$ \\
11*   & CH$_{3}$CCH           & 5$_{0}-4_{0}$                    & 85457   & 0.524   & 0.54    & $^{(3)}$ \\
14*   & H$^{13}$CO$^{+}$   & $1-0$                                   & 86754   & 4.351   & 3.97    &  $^{(12)}$ \\
14*   & H$^{13}$CO$^{+}$   & $1-0$                                   & 86754   & 4.351   & 4.56    & $^{(13)}$\\
14*   & SiO                               & $2-1$                                   & 86846   & 1.910   & 1.9      & $^{(4)}$ \\
14*   & SiO                               & $2-1$                                   & 86846   & 1.910   & 2.04    & $^{(12)}$ \\
14*   & SiO                               & $2-1$                                   & 86846   & 1.910   & 2.23    & $^{(14)}$ \\
14*   & SiO                               & $2-1$                                    & 86846   & 1.910   & 2.16    & $^{(15)}$ \\
19    & HCO$^{+}$                  & $1-0$                                    & 89188   & 15.195 & 13.93 & $^{(13)}$ \\
19    & CH$_{2}$DOH            & $2_{02}-1_{01}$ o$_{1}$ & 89251   & 0.182   & 0.19   & $^{(10)}$ \\
19    & CH$_{2}$DOH            & $2_{02}-1_{01}$ e$_{0}$ & 89407   & 0.262   & 0.31   & $^{(10)}$ \\
21    & CH$_{3}$OCHO-E     & $8_{08}-7_{07}$                & 90227   & 0.164   & 0.15   & $^{(3)}$ \\
21    & CH$_{3}$OCHO-A     & $8_{08}-7_{07}$                & 90229   & 0.144   & 0.11   & $^{(3)}$ \\
36    & $^{34}$SO                   & $3_{2}-2_{1}$                     & 97715   & 0.791   & 1         & $^{(4)}$ \\
38    & CH$_{3}$OCHO-A     & $8_{45}-7_{44}$                & 98682   & 0.248   & 0.22   & $^{(3)}$ \\
38    & CH$_{3}$OCHO-E     & $8_{45}-7_{44}$                & 98711   & 0.118   & 0.14   & $^{(3)}$ \\
60*  & C$^{18}$O                   & $1-0$                                    & 109782 & 9.900   & 9.02   & $^{(16)}$ \\
60*  & C$^{18}$O                   & $1-0$                                    & 109782 & 9.900   & 7.35   & $^{(17)}$\\
60*  & C$^{18}$O                   & $1-0$                                    & 109782 & 9.900   & 8.7     & $^{(13)}$ \\
61*  & CH$_{3}$CN               & $6_{40}-5_{40}$                 & 110349 & 0.164   & 0.48   & $^{(3)}$ \\
61*  & CH$_{3}$CN               & $6_{30}-5_{30}$                 & 110364 & 0.189   & 0.88   & $^{(3)}$ \\
61*  & CH$_{3}$CN               & $6_{20}-5_{20}$                 & 110375 & 0.304   & 0.77   & $^{(3)}$ \\
61*  & CH$_{3}$CN               & $6_{10}-5_{10}$ / $6_{00}-5_{00}$ & 110382  & 0.695 & 2.02& $^{(3)}$ \\
62   & D$_{2}$CO                   & $2_{12}-1_{11}$                 & 110838 & 0.515  & 0.64    & $^{(15)}$ \\
65   & C$^{17}$O                    & $1-0$                                    & 112359 & 2.855  & 3.07    & $^{(16)}$ \\
65   & C$^{17}$O                    & $1-0$                                    & 112359 & 2.855  & 2.59    & $^{(17)}$ \\
74   & SiO                                 & $3-2$                                    & 130268 & 1.361  & 4.86    & $^{(15)}$ \\
74   & SiO                                 & $3-2$                                    & 130268 & 1.361  & 4.61    & $^{(14)}$ \\
81   & CH$_{3}$CHO-E         & $7_{07}-6_{06}$                & 133830 & 0.383  & 0.41    & $^{(3)}$ \\
81   & CH$_{3}$CHO-A         & $7_{07}-6_{06}$                & 133854 & 0.597  & 0.57    & $^{(3)}$ \\
81   & CH$_{2}$DOH             & $3_{03}-2_{02}$ o$_{1}$ & 133872 & 0.739  & 0.6      & $^{(10)}$  \\
81   & CH$_{2}$DOH             & $3_{22}-2_{21}$ o$_{1}$ & 133881 & 0.465 & 0.34     & $^{(10)}$  \\
82   & HDCO                            & $2_{11}-1_{10}$                & 134284 & 1.466 & 1.8        & $^{(9)}$ \\
82   & HDCO                            & $2_{11}-1_{10}$                & 134284 & 1.466 & 1.55      &$^{(15)}$ \\
85   & SO$_{2}$                      & $5_{15}-4_{04}$                & 135696 & 3.115 & 4.6         & $^{(4)}$ \\
85   & SO$_{2}$                      & $5_{15}-4_{04}$                & 135696 & 3.115 & 3.41      & $^{(12)}$ \\
88   & H$_{2}^{13}$CO         & $2_{12}-1_{11}$                & 137450 & 0.524 & 0.58       & $^{(9)}$ \\
88   & H$_{2}^{13}$CO         & $2_{12}-1_{11}$                & 137450 & 0.524 & 0.58       & $^{(15)}$ \\
88   & H$_{2}^{13}$CO         & $2_{12}-1_{11}$                & 137450 & 0.524 & 0.76      & $^{(15)}$ \\
88   & H$_{2}^{13}$CO         & $2_{12}-1_{11}$                & 137450 & 0.524 & 0.76      & $^{(15)}$ \\
95   & H$_{2}$CO                  & $2_{12}-1_{11}$                & 140839 & 19.242 & 16.6    & $^{(9)}$ \\
95& H$_{2}$CO                     & $2_{12}-1_{11}$                & 140839 & 19.242 & 15.57  & $^{(15)}$ \\
100& C$_{2}$H$_{5}$CN    & 16$_{5,12}-15_{5,11}$ / 16$_{5,11}-15_{5,10}$ & 143406& 0.256& 0.55&  $^{(3)}$ \\
101& CH$_{3}$OCH$_{3}$ & $7_{3,4,2}-7_{2,5,2}$       & 143600 & 0.107   & 0.42     &  $^{(3)}$ \\
101& CH$_{3}$OCH$_{3}$ & $7_{3,4,1}-7_{2,5,1}$       & 143603 & 0.414   & 0.31     &  $^{(3)}$ \\
101& CH$_{3}$OCH$_{3}$ & $7_{3,4,0}-7_{2,5,0}$       & 143606 & 0.189   & 0.28     &  $^{(3)}$ \\
146& CHD$_{2}$OH             & $4_{0}-3_{0}$ e$_{0}$     & 166435 & 0.500   & 0.48      &  $^{(10)}$ \\
151& H$_{2}$S                      & $1_{10}-1_{01}$                & 168762 & 17.511 & 17.5     &  $^{(4)}$ \\
282*& H$_{2}^{13}$CO        & $3_{13}-2_{12}$                & 206132 & 1.752   & 1.6        &  $^{(9)}$ \\
282*& H$_{2}^{13}$CO        &$3_{13}-2_{12}$                 & 206132 & 1.752   & 1.97     & $^{(15)}$ \\
282*& H$_{2}^{13}$CO        & $3_{13}-2_{12}$                 & 206132 & 1.752   & 2.19     & $^{(15)}$ \\
282*& H$_{2}^{13}$CO        & $3_{13}-2_{12}$                 & 206132 & 1.752   & 2.48     & $^{(15)}$\\
282*& SO                                & $4_{5}-3_{4}$                      & 206176 & 21.367 & 36         &$^{(15)}$ \\
285*& CH$_{2}$DOH           & $2_{12}-3_{03}$ e$_{0}$ & 207780  & 0.690   & 1.01     & $^{(10)}$ \\
292*& H$_{2}$CO                 & $3_{13}-2_{12}$                 & 211211 & 31.416 & 43.8     & $^{(9)}$ \\
292*& H$_{2}$CO                 &  $3_{13}-2_{12}$                 & 211211 & 31.416 & 33        & $^{(15)}$ \\
292*& H$_{2}$CO                 &  $3_{13}-2_{12}$                 & 211211 & 31.416 & 26.4     &$^{(16)}$ \\
296& H$_{2}^{13}$CO         &  $3_{22}-2_{21}$                 & 213037 & 0.711    & 0.45    &$^{(15)}$ \\
296& H$_{2}^{13}$CO         &  $3_{21}-2_{20}$                 & 213294 & 0.593    & 0.46    &$^{(15)}$\\
303& H$_{2}$S                      &  $2_{20}-2_{11}$                 & 216710 & 7.436    & 3         & $^{(4)}$  \\
304& SiO                                 & $5-4$                                     & 217104 & 4.524    & 4.83    & $^{(14)}$ \\
312& D$_{2}$CO                   & $4_{14}-3_{13}$                 & 221191 & 1.274     & 1.6      & $^{(9)}$  \\
316& CH$_{2}$DOH             & $5_{23}-4_{14}$ e$_{1}$  & 223071 & 0.972     & 1.13   & $^{(10)}$ \\
316& CH$_{2}$DOH             & $5_{33}-4_{32}$ o$_{1}$  & 223153 & 1.204     & 0.58    & $^{(10)}$ \\
316& CH$_{2}$DOH             & $5_{23}-4_{22}$ e$_{1}$  & 223315 & 0.717     & 0.38    & $^{(10)}$ \\
321& HDO                               & $3_{12}-2_{21}$                 & 225897 & 2.089     & 1.7       & $^{(6)}$ \\
322& CH$_{3}$OCH$_{3}$ & $14Ð13$                               & 226346 & 1.685     & 0.89    & $^{(3)}$\\
323& CH$_{3}$CHO-E         & $12_{0,12}-11_{0,11}$      & 226551 & 0.809     & 0.49    & $^{(3)}$\\
323& CH$_{3}$OCHO-E      & $20_{2,18}-19_{2,18}$      & 226713 & 1.540     & 1.19    & $^{(3)}$\\
323& CH$_{3}$OCHO-A      & $20_{2,19}-19_{2,18}$      & 226718 & 0.707     & 0.49    & $^{(3)}$\\
323& CH$_{3}$OCHO-A      & $20_{1,19}-19_{1,18}$      & 226778 & 1.324     & 0.84    & $^{(3)}$\\
332& D$_{2}$CO                   & $4_{04}-3_{03}$                 & 231410 & 3.334     & 2.43    & $^{(15)}$ \\
332& D$_{2}$CO                   & $4_{04}-3_{03}$                 & 231410 & 3.334    & 2.01    & $^{(16)}$ \\
332& D$_{2}$CO                   & $4_{04}-3_{03}$                 & 231410 & 3.334    & 1.92    & $^{(17)}$ \\
332& D$_{2}$CO                   & $4_{04}-3_{03}$                 & 231410 & 3.334    & 2.69    &$^{(13)}$ \\
332& D$_{2}$CO                   & $4_{04}-3_{03}$                 & 231410 & 3.340    & 3          & $^{(9)}$ \\
332& CH$_{3}$CHO-A         & $12_{48}-11_{47}$            & 231457 & 0.520    & 0.51    & $^{(3)}$\\
342& D$_{2}$CO                   & $4_{22}-3_{21}$                 & 236102 & 1.718    & 1.86    & $^{(8)}$ \\
342& D$_{2}$CO                   & $4_{22}-3_{21}$                 & 236102 & 1.718    & 1.9      & $^{(9)}$ \\
342& D$_{2}$CO                   & $4_{22}-3_{21}$                 & 236102 & 1.710     & 1.9     & $^{(15)}$ \\
353& HDO                               &  $2_{11}-2_{12}$                & 241561 & 2.268     & 2        & $^{(6)}$ \\
363*& HDCO                          &  $4_{14}-3_{13}$                & 246925 & 2.067     & 8.43  & $^{(15)}$ \\
363*& HDCO                          &  $4_{14}-3_{13}$                & 246925 & 2.067     & 7.6    & $^{(9)}$ \\
401& HDO                               &  $2_{20}-3_{13}$               & 266161 & 0.214      & 0.21 & $^{(6)}$\\
434& C$^{18}$O                    & $3-2$                                   & 329335 & 55.768    & 33.6  & $^{(11)}$ \\
436& $^{13}$CH$_{3}$OH &  $7_{17}-6_{16}$                & 330194 & 0.458      & 0.51  & $^{(5)}$ \\
436& $^{13}$CH$_{3}$OH &  $7_{34}-6_{33}$                & 330408 & 1.123      & 0.91  & $^{(5)}$ \\
437& $^{13}$CO                   & $3-2$                                    & 330588 & 55.854    & 67.2  & $^{(11)}$ \\
446& HDCO                           &  $5_{14}-4_{13}$                & 335096 & 4.550      & 3.95  & $^{(2)}$ \\
446& HDCO                           &  $5_{14}-4_{13}$                & 335097 & 4.550      & 4.5    & $^{(9)}$ \\
450& C$^{17}$O                   & $3-2$                                    & 337061 & 12.481    & 11     & $^{(1)}$ \\
450& CH$_{3}$OH               & $3_{3}-4_{2}$  E$^{+}$     & 337136 & 0.826      & 1.05  & $^{(2)}$ \\
452& H$_{2}$CS                  & $21_{1,10}-9_{1,9}$          & 338081 & 1.723      & 2.4     & $^{(1)}$ \\
455& $^{34}$SO                   & $8_{9}-7_{8}$                     & 339858 & 3.863      & 2.98   & $^{(1)}$ \\
461& D$_{2}$CO                  & $6_{06}-5_{05}$                & 342522 & 1.875      & 1.9     & $^{(9)}$ \\
461& CS                                 & $7-6$                                    & 342883 & 38.535    & 51.4  & $^{(11)}$ \\
462& H$_{2}^{13}$CO        & $5_{15}-4_{14}$                 & 343325 & 3.181      & 1.37   & $^{(9)}$ \\
464& SO                                 & $8_{8}-7_{7}$                     & 344311 & 18.647    & 18.1   & $^{(1)}$ \\
470& SiO                                & $8-7$                                    & 347331 & 8.546      & 5.79   & $^{(1)}$ \\
475& D$_{2}$CO                 &  $6_{25}-5_{24}$                & 349631 & 1.506      & 1         & $^{(9)}$ \\
479& H$_{2}$CO                 &  $5_{15}-4_{14}$                & 351768 & 35.945    & 38.3   & $^{(9)}$  \\
483& H$_{2}^{13}$CO        &  $5_{05}-4_{04}$                & 353812 & 0.869      & 0.6     & $^{(9)}$  \\
484& HCN                             & $4-3$                                    & 354505 & 41.068    & 63.4   & $^{11)}$  \\
485& H$_{2}^{13}$CO        &  $5_{24}-4_{23}$               & 354899 & 0.557      & 0.3      & $^{(9)}$  \\
490& SO$_{2}$                     &  $15_{4,12}-15_{3,13}$   & 357241 & 3.603      & 3.29    & $^{(1)}$  \\
490& SO$_{2}$                     &  $11_{48}-11_{39}$          & 357388 & 4.201     & 2.25    & $^{(1)}$  \\
496& DCO$^{+}$                  & $5-4$                                   & 360170 & 3.135     & 4.02    & $^{(2)}$  \\
501& HNC                              & $4-3$                                   & 362630 & 8.264     & 11.9    & $^{(11)}$  \\
501& H$_{2}$CO                  &  $5_{05}-4_{04}$              & 362736  & 18.394   & 28.9   & $^{(9)}$  \\
503& H$_{2}$CO                  &  $5_{24}-4_{23}$              & 363946  & 11.728   & 11.9   & $^{(9)}$  \\
\end{longtable}
\noindent All observations are obtained with the same telescopes: IRAM\,30\,m from 80 to 280\,GHz, JCMT from 328 to 366\,GHz. The coordinates of the observed position are $\alpha$(2000.0)\,=\,16$^h$32$^m$ 22.6$^m$, $\delta$(2000.0)\,=\,$-$24$^{\circ}$ 28$'$ 33$''$; except for the settings numbers tagged with an asterisk that correspond to a slightly different position: $\alpha$(2000.0)\,=\,16$^h$32$^m$ 22.7$^m$, $\delta$(2000.0)\,=\,$-$24$^{\circ}$ 22$'$ 13$''$. References: $^{1}$\cite{1994ApJ...428..680B};  $^{2}$\cite{1995ApJ...447..760V};   $^{3}$\cite{2003ApJ...593L..51C}; $^{4}$\cite{2004A&A...413..609W};  $^{5}$\cite{2004A&A...416..159P};   $^{6}$\cite{2005A&A...431..547P};  $^{7}$\cite{2007ApJ...659L.137B}; $^{8}$\cite{1998A&A...338L..43C}; $^{9}$\cite{2000A&A...359.1169L};  $^{10}$\cite{2002A&A...393L..49P};  $^{11}$\cite{2002A&A...390.1001S},  unpublished IRAM\,30\,m observations: $^{12}$June 1997, $^{13}$September 2000, $^{14}$April 1998, $^{15}$January 1999, $^{16}$March 2000, $^{17}$August 2000, $^{18}$March 1998.}
} 
%
%
\onltab{3}{
\begin{table*}
\caption{Comparison of the strong spectral survey lines simultaneously observed with different backends.}
\label{tab_internal_calib}
\begin{center}
\begin{tabular}{rlrr|rlrr|rlrr}
\hline
\hline
Setting & Frequency &  \multicolumn{2}{c|}{$\int$  T$_A^*$ $d$v} & Setting & Frequency &  \multicolumn{2}{c|}{$\int$  T$_A^*$ $d$v} & Setting & Frequency &  \multicolumn{2}{c}{$\int$  T$_A^*$ $d$v}  \\
number & (MHz) &  \multicolumn{2}{c|}{(K.kms$^{-1}$)} & number & (MHz) &  \multicolumn{2}{c|}{(K.kms$^{-1}$)} &number & (MHz) &  \multicolumn{2}{c}{(K.kms$^{-1}$)} \\
&& VESPA &FB  &&& VESPA & FB  &&& VESPA & FB   \\
\hline
73 & 129515 & 2.436 & 3.091 & 105 & 145603 & 11.325 & 11.266 & 136 & 161172 & 0.501 & 0.367   \\
74 & 130268 & 1.361 & 5.378 & 106 & 146368 & 2.447 & 1.604 & 137 & 161603 & 0.462 & 0.713   \\
75 & 130892 & 0.506 & 0.290 & 107 & 146968 & 28.254 & 27.119 & 138 & 162186 & 0.340 & 0.302   \\
76 & 131015 & 1.432 & 2.062 & 108 & 147174 & 3.356 & 3.312 & 139 & 162530 & 0.779 & 0.900   \\
77 & 131885 & 1.103 & 1.042 & 109 & 147944 & 0.600 & 0.904 & 140 & 163119 & 1.409 & 1.142   \\
\hline
\end{tabular}
\end{center}
The two backends used in the 129-175\,GHz range were the VESPA autocorrelator (0.32\,MHz\,resolution) and the 1\,MHz resolution 
Filter Banks (FB). 
\end{table*}
} 
%
%
\onllongtabL{4}{
\small{
\begin{landscape}
\begin{longtable}{llrrlllrlrlrrl}
\caption[]{Identified non-blended lines. The line characteristics are derived from Gaussian fits.}
\label{lines-on-line}\\
\hline \hline\\[-1.8ex]
TAG & Species \& Transition & Frequency &  Eup & Aij & Vo        & $\delta_{Vo}$ & FWHM & $\delta_{FWHM}$ & Int  & $\delta_{Int}$  & Flux & $\delta_{Flux}$ & comments\\
         &                                        & (MHz)          &   (K)  & s$^{-1}$  &  \multicolumn{2}{c}{(kms$^{-1}$)} & \multicolumn{2}{c}{(kms$^{-1}$)}  & \multicolumn{2}{c}{(K)}  & \multicolumn{2}{c}{(K.kms$^{-1}$)}                           &  \\[0.5ex]
\hline\\[-1.8ex]
\endfirsthead
\caption{Continued.} \\
\hline\\[-1.8ex]
TAG & Species \& Transition & Frequency &  Eup & Aij & Vo        & $\delta_{Vo}$ & FWHM & $\delta_{FWHM}$ & Int  & $\delta_{Int}$  & Flux & $\delta_{Flux}$ & comments\\
         &                                        & (MHz)          &   (K)  & s$^{-1}$ &  \multicolumn{2}{c}{(kms$^{-1}$)} & \multicolumn{2}{c}{(kms$^{-1}$)}  & \multicolumn{2}{c}{(K)}  & \multicolumn{2}{c}{(K.kms$^{-1}$)}                       &  \\[0.5ex]
\hline\\[-1.8ex]
\endhead
\hline
\endfoot
\hline
\endlastfoot
28503   &   CO (1-0)      &115271.20   &5.53   &7.20E-08   &5.90   &0.23   &7.49   &0.54   &10.718   &0.667   &108.94   &11.98   &   broad, wings        \\   
28503   &   CO (2-1)      &230538.00   &16.60   &6.91E-07   &5.23   &0.39   &10.00   &0.92   &10.405   &0.826   &189.77   &32.26   &   broad, wings        \\   
28503   &   CO (3-2)      &345795.99   &33.19   &2.50E-06   &6.65   &0.40   &10.91   &0.94   &14.595   &1.087   &307.10   &55.28   &   broad, wings \\[0.5ex]  
29501   &   $^{13}$CO (1-0)      &110201.35   &5.29   &6.33E-08   &4.34   &0.06   &2.82   &0.14   &3.148   &0.135   &11.92 &1.31   &   small wings  \\   
29501   &   $^{13}$CO (2-1)      &220398.68   &15.87   &6.07E-07   &3.84   &0.10   &5.03   &0.22   &5.440   &0.210   &48.19 &8.19   &  small wings   \\   
29501   &   $^{13}$CO (3-2)      &330587.97   &31.73   &2.19E-06   &3.63   &0.47   &4.55   &1.13   &6.493   &1.377   &55.92 &10.07   &   wings  \\[0.5ex]  
30502   &   C$^{18}$O (1-0)      &109782.17   &5.27   &6.27E-08   &3.92   &0.03   &2.19   &0.06   &3.362   &0.081   &9.88   &1.09   &     \\   
30502   &   C$^{18}$O (2-1)      &219560.35   &15.81   &6.01E-07   &3.57   &0.04   &3.28   &0.10   &4.842   &0.131   &27.88   &4.74   &     \\
\end{longtable}
\noindent All lines have been identified with the line identification package CASSIS (http://cassis.cesr.fr), 
except the deuterated forms of methanol. For these latter species, not yet included in the JCMT and CDMS 
databases, the molecular data come from \cite{2002A&A...393L..49P, 2004A&A...416..159P} and we 
derived the line parameters with the GILDAS-CLASS package (http://www.iram.fr/IRAMFR/GILDAS).
$^*$ The deuterated forms of methanol, not being included in the CDMS or JPL spectroscopic databases, 
do not have TAG and A$_{ij}$. CASSIS ingest the CDMS and JPL spectroscopic databases, and the TAGs 
are those given by these databases to identify the molecules. CASSIS also ingest the VASTEL spectroscopic 
database (http://www.astro.caltech.edu/~vastel/CHIPPENDALES), in which the separation of the ortho and para
forms has been performed for some molecules listed in CDMS and JPL. For this database, the TAG are adapted 
from those of the molecules in the original database. 
The columns $\delta_{Int}$ and $\delta_{FWHM}$ only give respectively the statistical errors on the peak 
intensity and FWHM computed during the fit. The column $\delta_{Flux}$ gives the quadratic error on the 
total flux in the line, taking into account the calibration error given in the Section 3 Calibration. It is computed 
as: $\delta_{Flux}$ = $\sqrt{(Cal\, \times \,Flux)^2 + (rms^2\,\times 2\,\times \,FWHM\, \times \,dv)}$, 
where, respectively, Cal is the calibration error, Flux is the line flux, rms is the observed rms, FWHM is the 
full width half maximum of the line, and dv the velocity resolution, at the given line frequency.

\end{landscape}}
} 

\end{document}